\documentclass{svjour3}                     

\smartqed  
\usepackage{graphicx}

\begin{document}

\def\be{\begin{equation}}
\def\ee{\end{equation}}
\def\bea{\begin{eqnarray}}
\def\eea{\end{eqnarray}}
\def\orc{\Omega_{r_c}}
\def\om{\Omega_{m}}
\def\E{{\rm e}}
\def\bearst{\begin{eqnarray*}}
\def\eearst{\end{eqnarray*}}
\def\peleven{\parbox{11cm}}
\def\peffec{\peight{\bearst\eearst}\hfill\peleven}
\def\pspace{\peight{\bearst\eearst}\hfill}
\def\ptwelve{\parbox{12cm}}
\def\peight{\parbox{8mm}}

\markboth{M. S. Movahed, M.Farhang and S. Rahvar} {Observational
Constraints with recent Data on the DGP Modified Gravity}


\title{Observational Constraints with Recent Data on the DGP Modified Gravity}

\author{ M. Sadegh Movahed, Marzieh Farhang and Sohrab Rahvar}

\institute{M. Sadegh Movahed \at Department of Physics, Shahid
\\ Beheshti University, Evin, Tehran 19839, Iran
           \&
School of Astronomy, IPM (Institute for Studies in theoretical
Physics and Mathematics),\\ P.O.Box 19395-5531,Tehran, Iran
\and
           Marzieh Farhang \at
Department of Physics, Sharif University of Technology, P.O.Box
11365--9161, Tehran, Iran
\and
 Sohrab Rahvar \at
 Department of Physics, Sharif University of
Technology, P.O.Box 11365--9161, Tehran, Iran }


\maketitle

\begin{abstract}
We study one of the simplest covariant modified-gravity models based
on the Dvali-Gabadadze-Porrati (DGP) brane cosmology, a
self-accelerating universe. In this model gravitational leakage into
extra dimensions is responsible of late-time acceleration. We mainly
focus on the effects of the model parameters on the geometry and the
age of universe. Also we investigate the evolution of matter density
perturbations in the modified gravity model, and obtain an
analytical expression for the growth index, $f$. We show that
increasing $\Omega_{r_c}$ leads to less growth of the density
contrast $\delta$, and also decreases the growth index. We give a
fitting formula for the growth index at the present time and
indicate that dominant term in this expression verifies the
well-known approximation relation $f\simeq \Omega_m^{\gamma}$. As
the observational test, the new Supernova Type Ia (SNIa) Gold sample
and Supernova Legacy Survey (SNLS) data, size of baryonic acoustic
peak from Sloan Digital Sky Survey (SDSS), the position of the
acoustic peak from the CMB observations and the Cluster Baryon Gas
Mass Fraction (gas) are used to constrain the parameters of the DGP
model. We also combine previous results with large scale structure
formation (LSS) from the $2$dFGRS survey. Finally to check the
consistency of the DGP model, we compare the age of old cosmological
objects with age of universe in this model.
\end{abstract}
\section{Introduction}
Recent Observations of type Ia supernova (SNIa) provides the main
evidence for accelerating expansion of the Universe
\cite{ris,permul}. Analysis of SNIa and the Cosmic Microwave
Background radiation (CMB) observations indicates that about $70\%$
of the total energy of the Universe is made by the dark energy and
the rest of it is the dark matter with a few percent of Baryonic
matter \cite{bennett,peri,spe03}. The "cosmological constant" is a
possible explanation for the acceleration of the universe
\cite{wein}. This term in Einstein field equations can be regarded
as a fluid with the equation of state of $w=-1$. However, there are
two problems with the cosmological constant, namely the {\it
fine-tuning} and the {\it cosmic coincidence}. In the framework of
quantum field theory, the vacuum expectation value is $123$ order of
magnitude larger than the observed value of $10^{-47}$ GeV$^{4}$.
The absence of a fundamental mechanism which sets the cosmological
constant to zero or to a very small value is the cosmological
constant problem. The second problem known as the cosmic
coincidence, states that why are the energy densities of dark energy
and dark matter nearly equal today?

There are various solutions for this problem as the decays
cosmological constant models. A non-dissipative minimally coupled
scalar field, the so-called Quintessence model can play the role of
this time varying cosmological constant \cite{wet88,amen,peb88}. The
ratio of energy density of this field to the matter density in this
model increases by the expansion of the universe and after a while
dark energy becomes the dominant term of the energy-momentum tensor.
One of the features of this model is the variation of equation of
state during the expansion of the universe. Various Quintessence
models like k-essence \cite{arm00}, tachyonic matter \cite{pad03},
Phantom \cite{cal02,cal03} and Chaplygin gas \cite{kam01} provide
various equations of states for the dark energy
{\cite{cal03,kam01,wan00,per99,pag03,dor01,dor02,dor04,arb05}.

Another approach dealing with this problem is using the modified
gravity by changing the Einstein-Hilbert action. Some of models as
$1/R$ and logarithmic models provide an acceleration for the
universe at the present time \cite{bar05}. In addition to the
phenomenological modification of action, the brane cosmology also
implies modification for the general relativity on a brane embedded
in an extra dimension space. Some brane world models which produce
the late time acceleration have been tested using many observational
experiments such as local gravity \cite{Lue03,lue03,lue04},
Supernova Type Ia
\cite{bar05,zong,deff021,ane02,dab04,alm05,maartens1,saf}, angular
size of compact ratio sources \cite{alca02}, the age measurements of
high redshift objects \cite{alca021}, the optical gravitational
lensing surveys \cite{jain02}, the large scale structures
\cite{mul03}, and the X-ray gas mass fraction in galaxy clusters
\cite{zhu,alca05}.

In some recent papers \cite{maartens1,Barger,mf06} observational
constraints have been obtained through the old data of Supernova
Gold sample and its combination with CMB shift parameter and Baryon
acoustic oscillation. Recently Guo et al. have put constraints on
this model using recent SNIa data and Baryon acoustic oscillation
\cite{zong}. Song et al., Sawichi and Carroll have separately
investigated the effect of DGP on the integrated Sachs-Wolfe and
tested the validity of modified linear growth factor in the
sub-horizon scale \cite{yong,saw05}. In ref. \cite{yong} linear
growth of density contrast in this model has been reviewed but some
of the interesting quantity as the growth index and its behavior
versus redshift and its dependency to the model parameter is missed.

In this paper we examine the effects of DGP model on the geometrical
parameters of the universe. On the other hand we use the
observational results related to the background evolution. Since the
structure formation in DGP is currently well understood on scales
between a few percent of the Hubble scale and the scale radius of a
typical dark matter halo \cite{maartens2}, we combine those results
with the linear structure formation of large scale in the universe.
Meanwhile we concentrate our attention to the effect of
$\Omega_{r_c}$ and $\Omega_m$ as free parameters of the model on the
density contrast and growth index evolution. We extend the simplest
growth index analytic formula given for the flat $\Lambda$CDM
\cite{f} in the underlying modified gravity model. We organize this
paper as follows: In Sec.\ref{dgp1} we introduce DGP model as a
self-accelerating cosmology. Its free parameters and modified
Friedman equation which governs on the background dynamics of the
universe are also investigated. In Sec.\ref{lowredshift} we study
the effect of this model on the  comoving distance, comoving volume
element, the variation of angular size by the redshift\cite{alc79}.
In Sec. \ref{cobs} we put some constraints on the parameters of
model by using the background evolution, such as new Gold sample and
Legacy Survey of Supernova Type Ia data \cite{R04}, the position of
the observed acoustic angular scale on the last scattering surface,
CMB shift parameter, the baryonic oscillation length scale and
baryon gas mass fraction for the range of redshift, $z\leq1.0$. We
study the linear structure formation in this model and compare the
growth index with the observations from the $2-$degree Field Galaxy
Redshift Survey ($2$dFGRS) data in Sec. \ref{cstructure}. We also
compare the age of the universe in this model with the age of old
cosmological structures in Sec. \ref{agesection}. Sec.\ref{conc}
contains summary and conclusion of this work.

\section*{2. DGP MODIFIED GRAVITY}\label{dgp1}

One of the simplest covariant modified-gravity models is based on
the Dvali-Gabadadze-Porrati (DGP) brane-world model, as generalized
to cosmology by Deffayet~\cite{Dvali:2000rv}. (It is worth noting
that the original DGP model with a Minkowski brane was not
introduced to explain acceleration -- the generalization to a
Friedman brane was subsequently found to be self-accelerating.) In
this model, gravity leaks off the 4-dimensional brane universe into
the 5-dimensional bulk spacetime at large scales. Ordinary matter is
considered to be localized on the brane while gravity can propagate
in the bulk. At small scales, gravity is effectively bound to the
brane and 4D gravity is recovered to a good approximation. The
action for the five-dimensional theory is:

\begin{equation}
S=\frac{1}{2\kappa_5^2}\int
d^5x\sqrt{-g_{(5)}}R_{(5)}+\frac{1}{2\kappa_4^2}\int
d^4x\sqrt{-g_{(4)}}R_{(4)} + S_{\rm matter} \label{action1}
\end{equation}
where the subscripts $4$ and $5$ denote the quantities on the brane
and in the bulk, respectively, $\kappa_4^2(\kappa_5^2)$ is the
inverse of four(five)-dimensional reduced Planck mass, and $S_{\rm
matter}$ is the action for matter on the brane. The solution of DGP
action in FRW metric provides a self-accelerating universe for the
expanding phase of the universe. This model can be an alternative to
the cosmological constant for describing the present acceleration of
the universe. However, DGP model suffers from the ghost instability
that was shown in \cite{ghost1,ghost2} though the boundary effective
action formalism. On the other hand the existence of ghost is
confirmed by explicit calculation of the spectrum of linear
perturbations in the five-dimensional framework \cite{ghost3}. A
solution for this problem is so-called cascading DGP model in which
the unlike previous attempts, it is free of ghost instabilities. In
this model the 4D propagator is regulated by embedding the 3-brane
within a 4-brane with their own gravity terms induced by a flat 6D
bulk \cite{ghost4}.

Coming back to the action (\ref{action1}), at moderate scales the
induced gravity term is responsible for the recovery of
4-Dimensional Einstein gravity. The transition from 4D to 5D
behavior is governed by a crossover scale $r_c$:
\begin{equation}
r_c\equiv\frac{\kappa_5^2}{2\kappa_4^2}.
\end{equation}
In the weak-field gravitational field, potential behaves as $r^{-1}$
for $ r\ll r_c$ and as $r^{-2}$ for $ r\gg r_c$. At large scale
gravity is five dimensional. The energy conservation equation
remains the same as in general relativity, but the Friedman equation
is modified:
\begin{eqnarray}
&& \dot\rho+3H(\rho+p)=0\,,\label{ec} \\ && H^2+{K \over a^2}\pm{1
\over r_c}\sqrt{H^2+{K \over a^2}} = {8\pi G \over 3}\rho \label{f}
\end{eqnarray}
Two given sign in equation (\ref{f}) correspond to the two branches
of the cosmological evolution. The upper sing shows a de Sitter
expansion of the universe, while the lower sign corresponds to the
self-accelerating solution without the cosmological constant. So we
infer the cosmological effects of the second branch of DGP model.
Equations (\ref{ec}) and (\ref{f}) imply (for the CDM case $p=0$)
\begin{equation} \dot H- {K \over a^2}=-4\pi G\rho\left[ 1 + {1\over \sqrt{1+
32\pi G r_c^2\rho/3}} \right] \label{r}
\end{equation}
Equation~(\ref{f}) shows that at early times, when $H^2 +K/a^2 \gg
r_c^{-2}$, the general relativistic Friedman equation is recovered.
By contrast, at late times in a CDM universe, with $\rho\propto
a^{-3}\to 0$, we have
\begin{equation}
H\to H_\infty= {1\over r_c}
\end{equation}
Gravity leakage at late times initiates acceleration  not due to any
negative pressure field, but due to the weakening of gravity on the
brane. Since $H_0>H_\infty$, in order to achieve self-acceleration
at late times, we require
 \begin{equation}
r_c > H_0^{-1}
 \end{equation}
and this is confirmed by fitting observations as discussed below.

In dimensionless form, the modified Friedman equation~(\ref{f}) is
given by
 \begin{eqnarray}
{H(z)^2 \over H_0^2}&=&\left[ \sqrt{\om (1+z)^3+ \orc}+\sqrt{\orc}
\right]^2 \nonumber \\&&~~{}+\Omega_K(1+z)^2\,\label{fz}
 \end{eqnarray}
where
 \begin{eqnarray}
 \Omega_K &=& 1-\om-2\sqrt{\orc}\left(\!\sqrt{\orc}+\sqrt{\orc+\om}
\right)\nonumber\\
&=&1-\Omega_{tot}\!\! \label{ok}\\
\orc &=& {1 \over 4 H_0^2 r_c^2}\,.
 \end{eqnarray}
From equation~(\ref{r}), the dimensionless acceleration is
 \begin{eqnarray}
q&=&{1\over H_0^2}{\ddot a \over a}=\left( \sqrt{\om (1+z)^3+
\orc}+\sqrt{\orc}\right)\Biggl[ \sqrt{\orc} \nonumber\\ &&~{}+
{2\orc-\om(1+z)^3 \over 2\sqrt{\om (1+z)^3+ \orc} }\Biggr]
 \end{eqnarray}
so that the redshift at which acceleration era is started is given
by \cite{zhu}
 \begin{equation}
 z_{q=0}=2\left( {\orc \over \om} \right)^{\frac{1}{3}}-1 \label{za}
 \end{equation}
 Also equation (\ref{ok}) shows in the flat model:
 \begin{equation}
\orc ={1 \over 4}(1-\om)^2
 \end{equation}
For the transition from negative to the positive acceleration at the
present-time, equation (\ref{za}) implies :
 \begin{equation}
\orc={\om \over 8}
 \end{equation}
Acceleration parameter for this model in terms of scale factor is
shown in Figure \ref{acc}. Increasing the value of $\Omega_{r_{c}}$
causes that universe entered in the acceleration phase at the
earlier times. The lower panel of Figure \ref{acc} shows
acceleration parameter for the flat $\Lambda$CDM model, obviously
$\Omega_{r_c}$ has the same role as cosmological constant.

The modified Friedman equation in DGP may be reinterpreted from a
standard viewpoint. We define the effective dark energy density
$\rho_{eff}\equiv 3H/8\pi Gr_c$. Then the effective dark energy
equation of state $w_{eff}\equiv p_{eff}/\rho_{eff}$ is given by
$\dot{\rho}_{eff} +3H(1+w_{eff}) \rho_{eff}=0$. Thus $\rho_{eff}$
and $w_{eff}$ give a standard general relativistic interpretation of
DGP expansion history, i.e., they describe the equivalent general
relativity dark energy model. For the flat case, $\Omega_K=0$, we
find
 \begin{equation} \label{wx}
w_{eff}(z)={\om-1-\sqrt{(1-\om)^2+4\om(1+z)^3} \over
2\sqrt{(1-\om)^2+4\om(1+z)^3}}\,,
 \end{equation}
which implies
 \begin{equation}\label{wx0}
w_{eff}(0)=-{1\over 1+\om}\,.
 \end{equation}

The DGP and $\Lambda$CDM models have the same number of parameters,
with $r_c$ substituting  $\Lambda$, therefore DGP model gives a very
useful framework for comparing the $\Lambda$CDM general relativistic
cosmology to a modified gravity alternative. Now an interesting
question that arises is: "can DGP model predict dynamics of
universe?" or in another word, "what values of the model parameter
to be consistent with observational tests?"

In the forthcoming sections we will study the observational
constraints on the model.

\begin{figure}[t]
\centerline{\includegraphics[width=0.7\textwidth]{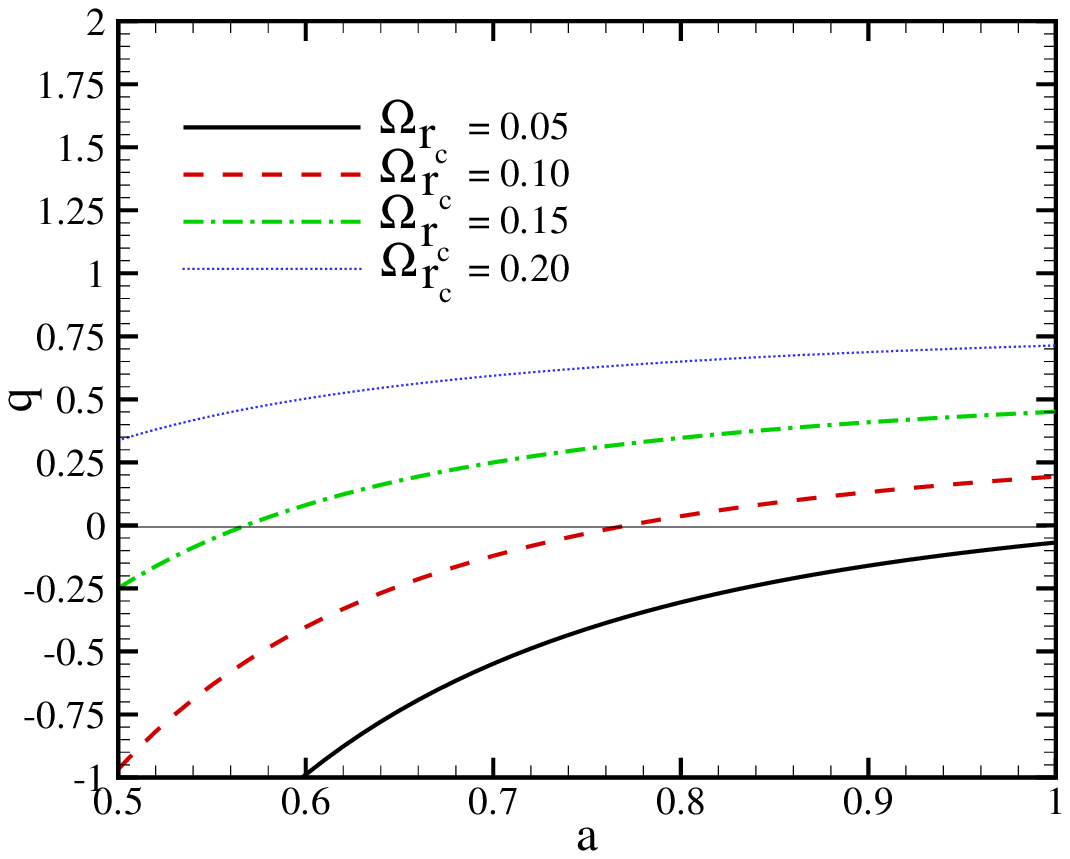}}
\centerline{\includegraphics[width=0.7\textwidth]{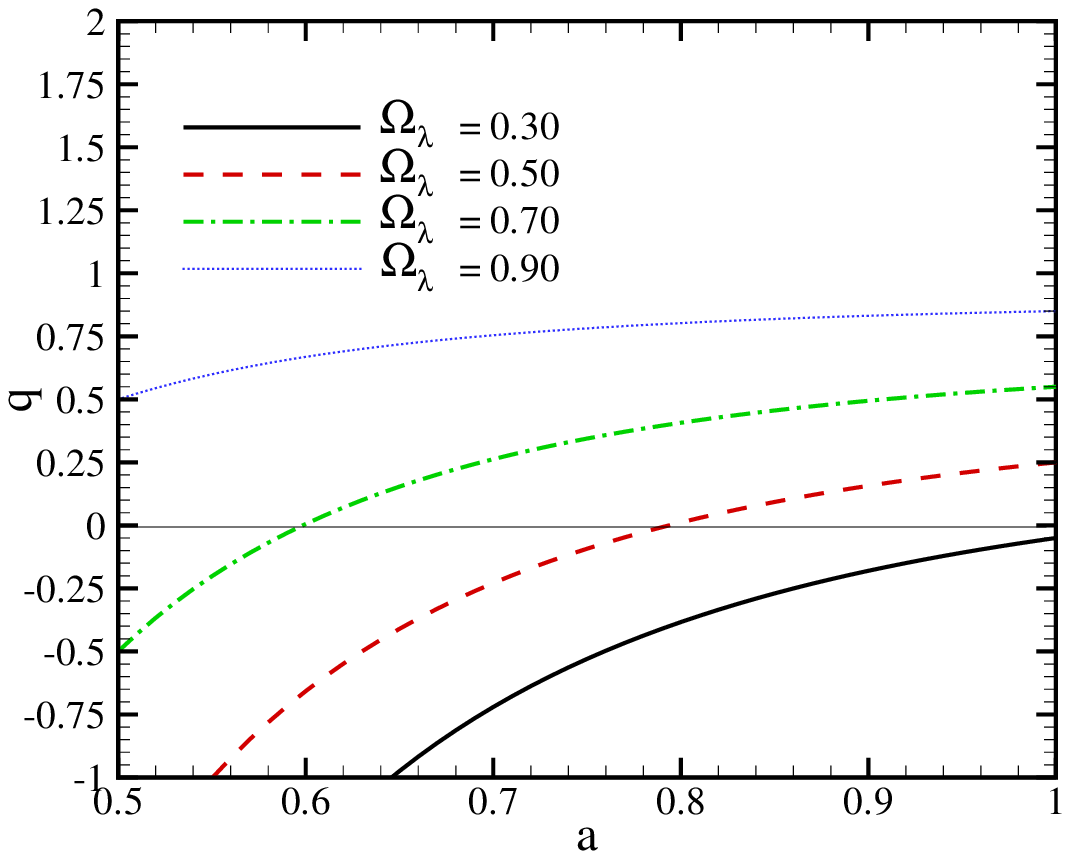}}
\caption{ Upper panel shows acceleration parameter
$(q=\ddot{a}/{H_0^2a})$ in the DGP model as a function of scale
factor for various values of $\Omega_{r_{c}}$. Lower panel
corresponds to the same function for the flat $\Lambda$CDM. We chose
the flat universe. } \label{acc}
\end{figure}

\section{The effect of DGP model on the geometrical parameters of universe}
\label{lowredshift} The cosmological observations are mainly
affected by the background dynamics of universe.
In this part we study the sensitivity of the geometrical parameters
on the parameters of DGP model.
\subsection{comoving distance}
The radial comoving distance is one of the basic parameters in
cosmology. For an object with the redshift of $z$, using the null
geodesics in the FRW metric, the comoving distance is obtained as:
\begin{eqnarray} r(z;\Omega_m,\Omega_{r_{c}}) &=& {1 \over
H_0\sqrt{|\Omega_K|}}\, {\cal F} \left( \sqrt{|\Omega_K|}\int_0^z\,
{dz' \over H(z')/H_0} \right), \label{comoving}\nonumber\\
 \end{eqnarray}
where
 \begin{eqnarray}
 {\cal F}(x) &\equiv & (x,\sin x, \sinh
x)~for~K=(0,1,-1)\,.
 \end{eqnarray}
and $H(z;\Omega_m,\Omega_{r_{c}})$ is given by equation (\ref{fz}).
\begin{figure}[t]
\centerline{\includegraphics[width=0.7\textwidth]{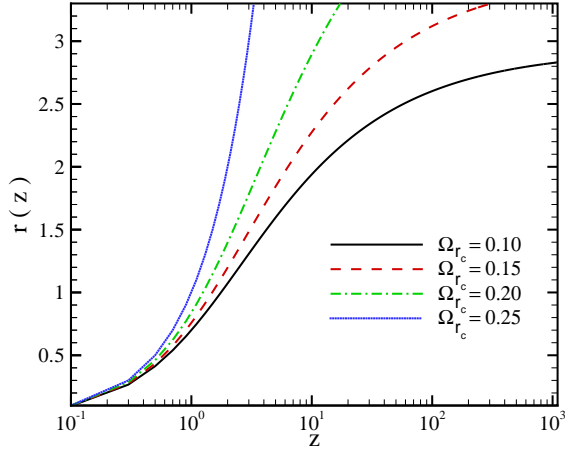}}
\caption{Comoving distance, $r(z;\Omega_m,\Omega_{r_{c}})$ (in unit
of $c/H_0$) as a function of redshift for various values of
$\Omega_{r_{c}}$. We fixed $\Omega_K= 0.0$.} \label{fig:rz}
 \end{figure}
By numerical integration of equation (\ref{comoving}), the comoving
distance in terms of redshift for different values of
$\Omega_{r_{c}}$ is shown in Figure~\ref{fig:rz}. Increasing the
$\Omega_{r_{c}}$ results in a longer comoving distance. According to
this behavior, by tuning the value of $\Omega_{r_c}$ we may explain
the Supernova Type Ia observations.
\subsection{Angular Size}
The apparent angular size of an object located at the cosmological
distance is another important parameter that can be affected by
the cosmological model during the history of the universe. An
object with the physical size of $D$ is related to the apparent
angular size of $\theta$ by:
\begin{equation}
D=d_A \theta \label{as}
\end{equation}
where $d_A=r(z;\Omega_m,\Omega_{r_{c}})/(1+z)$ is the angular
diameter distance. The main applications of equation (\ref{as}) is
on the measurement of the apparent angular size of acoustic peak
on CMB and baryonic acoustic peak at the high and low redshifts,
respectively. By measuring the angular size of an object in
different redshifts (the so-called Alcock-Paczynski test) it is
possible to probe the validity of modified gravity models
\cite{alc79}. The variation of apparent angular size
$\Delta\theta$ in terms of $\Delta z$ is given by:
\begin{equation}
{\Delta z\over \Delta \theta} =
H(z;\Omega_m,\Omega_{r_{c}})r(z;\Omega_m,\Omega_{r_{c}})
\label{alpa}
\end{equation}
\begin{figure}[t]
\centerline{\includegraphics[width=0.7\textwidth]{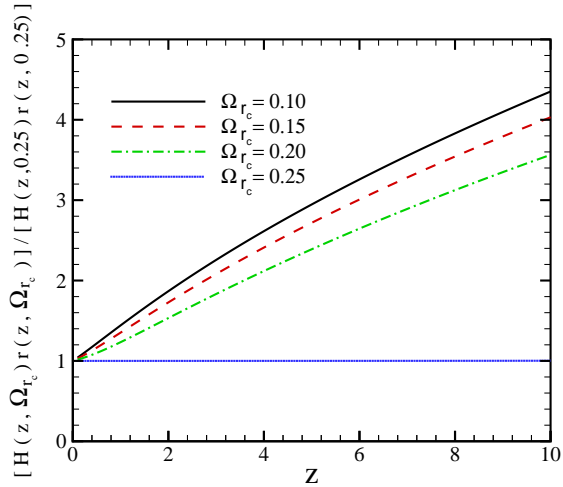}}
\caption{Alcock-Paczynski test comparing $\Delta z/{\Delta \theta}$
as a function of redshift for four different $\Omega_{r_{c}}$
normalized to the case with $\Omega_m=0$ and $\Omega_{r_{c}}=0.25$
(flat universe $\Omega_K=0$). It must be pointed out that for other
values of $\Omega_{r_{c}}$ we also assumed $\Omega_K=0.0$. }
\label{fig:hr}
 \end{figure}

Figure~\ref{fig:hr} shows $\Delta z/ \Delta \theta$ in terms of
redshift, normalized to the case with $\Omega_{m}=0.0$ and flat
universe $\Omega_K=0.0$. The advantage of Alcock-Paczynski test is
that it is independent of standard candles and knowing a standard
ruler such as the size of baryonic acoustic peak one can use it to
constrain the modified gravity model.
\subsection{Comoving Volume Element}
The comoving volume element is another geometrical parameter
which is used in number-count tests such as lensed quasars,
galaxies, or clusters of galaxies. The comoving volume element in
terms of comoving distance and Hubble parameter is given by:
\begin{equation}
f(z;\Omega_m,\Omega_{r_{c}}) \equiv {dV\over dz d\Omega} =
r^2(z;\Omega_m,\Omega_{r_{c}})/H(z;\Omega_m,\Omega_{r_{c}}).
\end{equation}
According to Figure \ref{fig:v}, the comoving volume element becomes
large for larger value of $\Omega_{r_{c}}$ in the flat universe.

\begin{figure}[t]
\centerline{\includegraphics[width=0.7\textwidth]{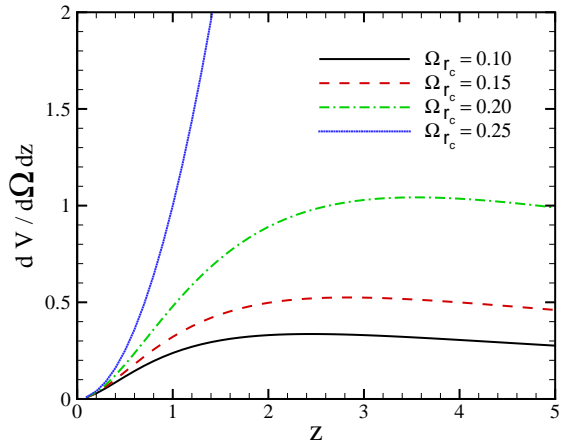}}
\caption{The comoving volume element in terms of redshift for
various $\Omega_{r_{c}}$ exponent. Increasing $\Omega_{r_{c}}$
shifts the position of the maximum value of volume element to lower
redshifts. We fixed $\Omega_K=0.0$}. \label{fig:v}
 \end{figure}

\section{Observational Constraints From the Background Evolution}
\label{cobs}
 In this section we compare the SNIa Gold sample released from
 more recent observations \cite{new} which have lower systematic errors than
 the former Gold sample data set. This new catalog contains $156$ Supernova type Ia.
 On the other hand we also take into account $116$ supernova Legacy Survey data which seems to be more
  consistent with WMAP observation as another SNIa observations to examine DGP model. To make the acceptance interval of
  model free parameters more confined, we use the location of acoustic peak of temperature
  fluctuations from WMAP observation, the location of baryonic acoustic oscillation peak from the
  SDSS and baryon gas mass fraction in the cluster for $26$ samples at redshifts less than
  $1$ \cite{allen04}.
\label{sn} The Supernova Type Ia experiments provided the main
evidence of the existence of dark energy. Since 1995 two teams of
the {\it High-Z Supernova Search} and the {\it Supernova Cosmology
Project} have discovered several type Ia supernovas at the high
redshifts \cite{per99,Schmidt}. Recently Riess et al.(2004)
announced the discovery of $16$ type Ia supernova with the Hubble
Space Telescope. This new sample includes $6$ of the $7$ most
distant ($z> 1.25$) type Ia supernovas. They determined the
luminosity distance of these supernovas and with the previously
reported algorithms, obtained a uniform $156$ Gold sample of type Ia
supernovas as a new data set with lower systematic errors than
former Gold sample data \cite{R04,Tonry,bar04}. At the beginning we
compare the predictions of the DGP model with the recent SNIa Gold
sample \cite{new}. The observations of supernova measure essentially
the apparent magnitude $m$ including reddening, K correction, etc,
which are related to the (dimensionless) luminosity distance, $D_L$,
of an object at redshift $z$ through:
\begin{equation}
m={\mathcal{M}}+5\log{D_{L}(z;\Omega_m,\Omega_{r_{c}})}, \label{m}
\end{equation} where
\begin{eqnarray}
\label{luminosity} D_L (z;\Omega_m,\Omega_{r_{c}}) &=&{(1+z) \over
\sqrt{|\Omega_K|}}\, {\cal F} \left( \sqrt{|\Omega_K|}\int_0^z\,
{dz'H_0\over H(z')} \right).
\end{eqnarray}
Also
\begin{eqnarray}
\label{m1}\mathcal{M} &=& M+5\log{\left(\frac{c/H_0}{1\quad
Mpc}\right)}+25.
\end{eqnarray}
where $M$ is the absolute magnitude. The distance modulus, $\mu$, is
defined as:

\begin{eqnarray}
\mu\equiv
m-M&=&5\log{D_{L}(z;\Omega_m,\Omega_{r_{c}})}\nonumber\\&&\quad
+5\log{\left(\frac{c/H_0}{1\quad Mpc}\right)}+25, \label{eq:mMr}
\end{eqnarray}
or

\begin{eqnarray}
\mu&=&5\log{D_{L}(z;\Omega_m,\Omega_{r_{c}})}+\bar{M}
\end{eqnarray}

In order to compare the theoretical results with the observational
data, we must compute the distance modulus, as given by equation
(\ref{eq:mMr}). For this purpose,the first step is to compute the
quality of the fitting through the least squared fitting quantity
$\chi^2$ defined by:
\begin{eqnarray}\label{chi_sn}
\chi^2(\bar{M},\Omega_m,\Omega_{r_{c}})&=&\sum_{i}\frac{[\mu_{obs}(z_i)-\mu_{th}(z_i;\Omega_m,\Omega_{r_{c}},\bar{M})]^2}{\sigma_i^2},\nonumber\\
\end{eqnarray}
where $\sigma_i$ is the observational uncertainty in the distance
modulus. To constrain the parameters of model, we use the Likelihood
statistical analysis:
\begin{eqnarray}\label{lik}
{\cal
L}(\bar{M},\Omega_m,\Omega_{r_{c}})={\mathcal{N}}e^{-\chi^2(\bar{M},\Omega_m,\Omega_{r_{c}})/2}
\end{eqnarray}
where ${\mathcal{N}}$ is a normalization factor. The parameter
$\bar{M}$ is a nuisance parameter and should be marginalized
(integrated out) leading to a new $\bar{\chi}^2$ defined as:
\begin{eqnarray}\label{mar2}
\bar{\chi}^2=-2\ln\int_{-\infty}^{+\infty}{\cal
L}(\bar{M},\Omega_m,\Omega_{r_{c}})d\bar{M}
\end{eqnarray}
 Using equations (\ref{chi_sn}), (\ref{lik}) and (\ref{mar2}), we find:
\begin{eqnarray}\label{mar3}
\bar{\chi}^2(\Omega_m,\Omega_{r_{c}})&=&\chi^2(\bar{M}=0,\Omega_m,\Omega_{r_{c}})-\frac{B(\Omega_m,\Omega_{r_{c}})^2}{C}\nonumber\\
&&+\ln(C/2\pi)
\end{eqnarray}
where
\begin{eqnarray}\label{mar4}
B(\Omega_m,\Omega_{r_{c}})=\sum_{i}\frac{[\mu_{obs}(z_i)-\mu_{th}(z_i;\Omega_m,\Omega_{r_{c}},\bar{M}=0)]}{\sigma_i^2},
\end{eqnarray}
\begin{eqnarray}\label{mar5}
C=\sum_{i}\frac{1}{\sigma_i^2},
\end{eqnarray}

Equivalent to marginalization is the minimization with respect to
$\bar{M}$. One can show that $\chi^2$ can be expanded in $\bar{M}$
as \cite{Nesseris04}:
\begin{eqnarray}\label{mar6}
\chi^2(\Omega_m,\Omega_{r_{c}})=\chi^2(\bar{M}=0,\Omega_m,\Omega_{r_{c}})-2\bar{M}B+\bar{M}^2C
\end{eqnarray}
which has a minimum for $\bar{M}=B/C$:
\begin{eqnarray}\label{mar7}
\chi^2_{\rm
SNIa}(\Omega_m,\Omega_{r_{c}})=\chi^2(\bar{M}=0,\Omega_m,\Omega_{r_{c}})-\frac{B(\Omega_m,\Omega_{r_{c}})^2}{C}
\end{eqnarray}

Using equation (\ref{mar7}) we can find the best fit values of model
parameters as the values that minimize
$\chi^2(\Omega_m,\Omega_{r_{c}})$. For the Likelihood analysis we
use some  weak priors for the model parameters indicated in Table
\ref{prior}.
\begin{figure}[t]
\centerline{\includegraphics[width=0.7\textwidth]{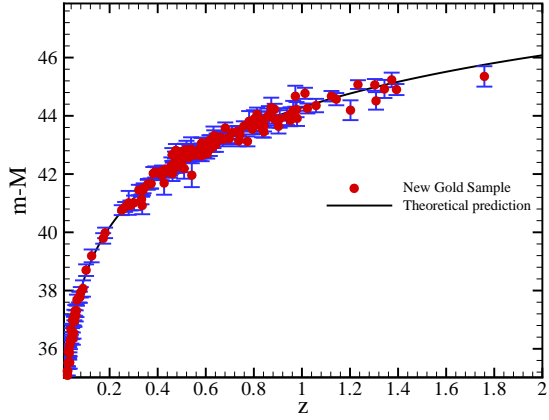}}
\caption{Distance modulus of the SNIa new Gold sample in terms of
redshift. Solid line shows the best fit values with the
corresponding parameters of $h=0.64$,
$\Omega_m=0.36^{+0.07}_{-0.06}$,
$\Omega_{r_{c}}=0.23_{-0.04}^{+0.04}$ in $1 \sigma$ level of
confidence with $\chi^2_{min}/N_{d.o.f} =0.91$ for DGP model.}
\label{modul1}
\end{figure}
\begin{figure}[t]
\centerline{\includegraphics[width=0.7\textwidth]{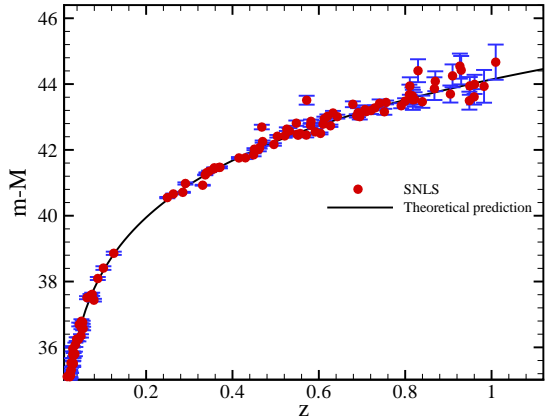}}
\caption{Distance modulus of the SNLS supernova data in terms of
redshift. Solid line shows the best fit values with the
corresponding parameters of $h=0.70$,
$\Omega_m=0.13^{+0.06}_{-0.06}$,
$\Omega_{r_{c}}=0.14_{-0.03}^{+0.03}$ in $1 \sigma$ level of
confidence with $\chi^2_{min}/N_{d.o.f} =0.85$ for DGP model.}
\label{modul2}
\end{figure}
\begin{table}
\begin{center}
\caption{Priors on the parameter space, used in the likelihood
analysis.} {\begin{tabular}{|ccc|}
  \hline
{\rm Parameter}& Prior & \\  \hline
  $\Omega_{tot}$ & - & {\rm Free}\\\hline
 $\Omega_m$& $0.00-1.00$& {\rm Top hat}\\\hline
 $\Omega_{r_c}$& $0.00-1.00$& {\rm Top hat}\\\hline
 $\Omega_bh^2$&$0.020\pm0.005$&{\rm Top hat (BBN)}
 \\ \hline
 $h$&$-$&
 \\
 \hline
 \end{tabular}
\label{prior}
}
\end{center}
\end{table}
The best values for the parameters of the model are:
$\Omega_m=0.36^{+0.07}_{-0.06}$,
$\Omega_{r_{c}}=0.23_{-0.04}^{+0.04}$ and
$\Omega_K=-0.56^{+0.20}_{-0.20}$ with $\chi^2_{min}/N_{d.o.f} =0.91$
at $1 \sigma$ level of confidence. The corresponding value for the
Hubble parameter at the minimized $\chi^2$ is $h=0.64$ and since we
have already marginalized over this parameter we do not assign an
error bar for it. The best fit values for the parameters of model by
using SNLS supernova data are $\Omega_m=0.13^{+0.06}_{-0.06}$,
$\Omega_{r_{c}}=0.14_{-0.03}^{+0.03}$ and
$\Omega_K=0.20^{+0.16}_{-0.16}$ with $\chi^2_{min}/N_{d.o.f} =0.85$
at $1 \sigma$ level of confidence. The value of Hubble parameter at
the minimum value of $\chi^2$ is $h=0.70$. Obviously our results are
different from what reported in \cite{zong} and \cite{maartens1}. In
the first reference they report the following values for the model
parameters: $\Omega_m=0.34^{+0.07}_{-0.08}$ and
$\Omega_{r_c}=0.24^{+0.04}_{-0.04}$ using old SNIa Gold sample while
in the second reference Marteens et. al. reported: $\Omega_m=0.270$
and $\Omega_{r_c}=0.125$. We point out that the constraint by SNIa
is very sensitive to the various catalogs of supernova data set
\cite{nesseris}. Table \ref{tab4} indicates the results from
observational constraints on the free parameters. Figures
\ref{modul1} and \ref{modul2} show the comparison of the theoretical
prediction of distance modulus by using the best fit values of model
parameters and observational values from new Gold sample and SNLS
supernova, respectively. For the age consistency test we substitute
the parameters of model from the SNIa new Gold sample and SNLS
fitting in equation (\ref{age}) (see Sec. \ref{agesection} for more
details) and obtain the age of universe about
$13.78_{-0.59}^{+0.68}$ Gry and $14.96_{-1.43}^{+1.03}$ Gry,
respectively. They give a universe older than what is expected from
the old stars.

%

The other constrain results from the CMB acoustic peak observations.
Before last scattering, the photons and baryons are tightly coupled
by Compton scattering and behave as a fluid. The oscillations of
this fluid, occurring as a result of the balance between the
gravitational interactions and the photon pressure, lead to the
familiar spectrum of peaks and troughs in the averaged temperature
anisotropy spectrum which we measure today. The odd and even peaks
correspond to maximum compression of the fluid and to rarefaction,
respectively \cite{hu96}. In an idealized model of the fluid, there
is an analytic relation for the location of the $m$-th peak: $l_m
\approx ml_A$ \cite{Hu95,hu00} where $l_A$ is the acoustic scale
which may be calculated analytically and depends on both pre- and
post-recombination physics as well as the geometry of the universe.
The acoustic scale corresponds to the Jeans length of photon-baryon
structures at the last scattering surface some $\sim 379$ Kyr after
the Big Bang \cite{spe03}. The apparent angular size of acoustic
peak can be obtained by dividing the comoving size of sound horizon
at the decoupling epoch $r_s(z_{dec})$ by the comoving distance of
observer to the last scattering surface $r(z_{dec})$:
\begin{equation}
\theta_A =\frac{\pi}{l_A}\equiv {{r_s(z_{dec})}\over r(z_{dec}) }.
\label{eq:theta_s}
\end{equation}
The size of sound horizon at the numerator of equation
(\ref{eq:theta_s}) corresponds to the distance that a perturbation
of pressure can travel from the beginning of universe up to the last
scattering surface and is given by:
 \bea
&&r_{s}(z_{dec};\Omega_m,\Omega_{r_c}) \nonumber\\
&&= {1 \over H_0\sqrt{|\Omega_k|}}\times {\cal F} \left(
\sqrt{|\Omega_k|}\int_{z_{dec}}^ {\infty} {v_s(z')dz' \over
H(z')/H_0} \right) \label{sh}
 \eea
where $v_s(z)^{-2}=3 + 9/4\times\rho_b(z)/\rho_{rad}(z)$ is the
sound velocity in the unit of speed of light from the big bang up to
the last scattering surface \cite{dor01,Hu95} and the redshift of
the last scattering surface, $z_{dec}$, is given by \cite{Hu95}:
\begin{eqnarray}\label{dec}
z_{dec} &=& 1048\left[ 1 + 0.00124(\omega_b)^{-0.738}\right]\left[
1+g_1(\omega_m)^{g_2}\right],\nonumber \\
g_1 &=& 0.0783(\omega_b)^{-0.238}\left[1+39.5(\omega_b)^{0.763}\right]^{-1},\nonumber \\
g_2 &=& 0.560\left[1 + 21.1(\omega_b)^{1.81}\right]^{-1},
\end{eqnarray}
where $\omega_m\equiv\Omega_mh^2$, $\omega_b\equiv\Omega_bh^2$ and
$\rho_{rad}$ is the radiation density. $\Omega_b$ is relative
baryonic density to the critical density at the present time.
Changing the parameters of the model can change the size of apparent
acoustic peak and subsequently the position of $l_A\equiv
\pi/\theta_A$ in the power spectrum of temperature fluctuations at
the last scattering surface. The simple relation $l_m\approx ml_A$
however does not hold very well for the peaks although it is better
for higher peaks \cite{hu00,doran01}. Driving effects from the decay
of the gravitational potential as well as contributions from the
Doppler shift of the oscillating fluid introduce a shift in the
spectrum. A good parametrization for the location of the peaks and
troughs is given by \cite{hu00,doran01}
\begin{equation}\label{phase shift}
l_m=l_A(m-\phi_m)
\end{equation}
where $\phi_m$ is phase shift determined predominantly by
pre-recombination physics, and are independent of the geometry of
the Universe. The location of acoustic peaks can be determined in
model by equation (\ref{phase shift}) with
$\phi_m(\omega_m,\omega_b)$. Doran et. al. \cite{doran01}, recently
have shown that the first and third phase shifts are approximately
model independent. The values of these shift parameters have been
reported as: $\phi_1(\omega_m,\omega_b)\simeq 0.27$ and
$\phi_3(\omega_m,\omega_b)\simeq 0.341$ \cite{hu00,doran01}.
According to the WMAP observations: $l_1=220.1\pm0.8$ and
$l_3=809\pm7$, so the corresponding observational values of
$l_A^{\rm obs}$ read as:
\begin{eqnarray}\label{peak1}
l_A^{\rm obs}|_{l_1}&=&\frac{l_1}{(1-\phi_1)}=299.45\pm2.67\\
l_A^{\rm obs}|_{l_3}&=&\frac{l_3}{(3-\phi_3)}=304.24\pm2.63
\end{eqnarray}
their Likelihood statistics are as follows:
 \begin{equation}\label{chil1}
  \chi_{l_1}^2=\frac{\left[l_A^{\rm obs}|_{l_1}-l_A^{\rm
th}|_{l_1}\right]^2}{\sigma_1^2}
\end{equation}
and
\begin{equation}\label{chil3}
  \chi_{l_3}^2=\frac{\left[l_A^{\rm obs}|_{l_3}-l_A^{\rm
th}|_{l_3}\right]^2}{\sigma_3^2}
\end{equation}
because of weak dependency of phase shift to the cosmological model
one can use another model independent parameter which is so-called
shift parameter ${\cal R}$:
\begin{equation}
{\cal R}\propto\frac{l_1^{flat}}{l_1},
\end{equation}
where $l_1^{flat}$ corresponds to the flat pure-CDM model with
$\Omega_m=1.0$ and the same $\omega_m$ and $\omega_b$ as the
original model.  It is easily shown that shift parameter is as
follows \cite{bond97}:
\begin{equation}\label{shift_th}
\label{shift} {\cal R}=
\sqrt{\Omega_m}\frac{D_L(z_{dec},\Omega_m,\Omega_{r_c})}{(1+z_{dec})}
\end{equation}
The observational results of CMB experiments correspond to a shift
parameter of ${\cal R}=1.716\pm0.062$ (given by WMAP, CBI, ACBAR)
\cite{spe03,pearson03}. One of the advantages of using the parameter
${\cal R}$ is that it is independent of Hubble constant. In order to
put constraint on the model from CMB, we compare the observed shift
parameter with that of model using likelihood statistic as
\cite{bond97}:
\begin{equation}
{\cal{L}}\sim e^{-\chi_{\rm CMB}^2/2}
\end{equation}
where \begin{equation}\label{chi_cmb} \chi_{\rm
CMB}^2=\frac{\left[{\cal R}_{{\rm obs}}-{\cal R}_{{\rm
th}}\right]^2}{\sigma_{\rm CMB}^2}
\end{equation}
where ${\cal R}_{{\rm th}}$ and ${\cal R}_{{\rm obs}}$ are
determined using equation (\ref{shift_th}) and given by observation,
respectively. Figures \ref{l1} shows $\Omega_{r_c}$ and
$\Omega_{\Lambda}$ in DGP model and the $\Lambda$CDM as a function
of $\Omega_m$ for a given $l_A$. Decreasing both $\Omega_{r_c}$ and
$\Omega_{\lambda}$ lead an increasing in the value of present matter
density.

\begin{figure}[t]
\centerline{\includegraphics[width=0.7\textwidth]{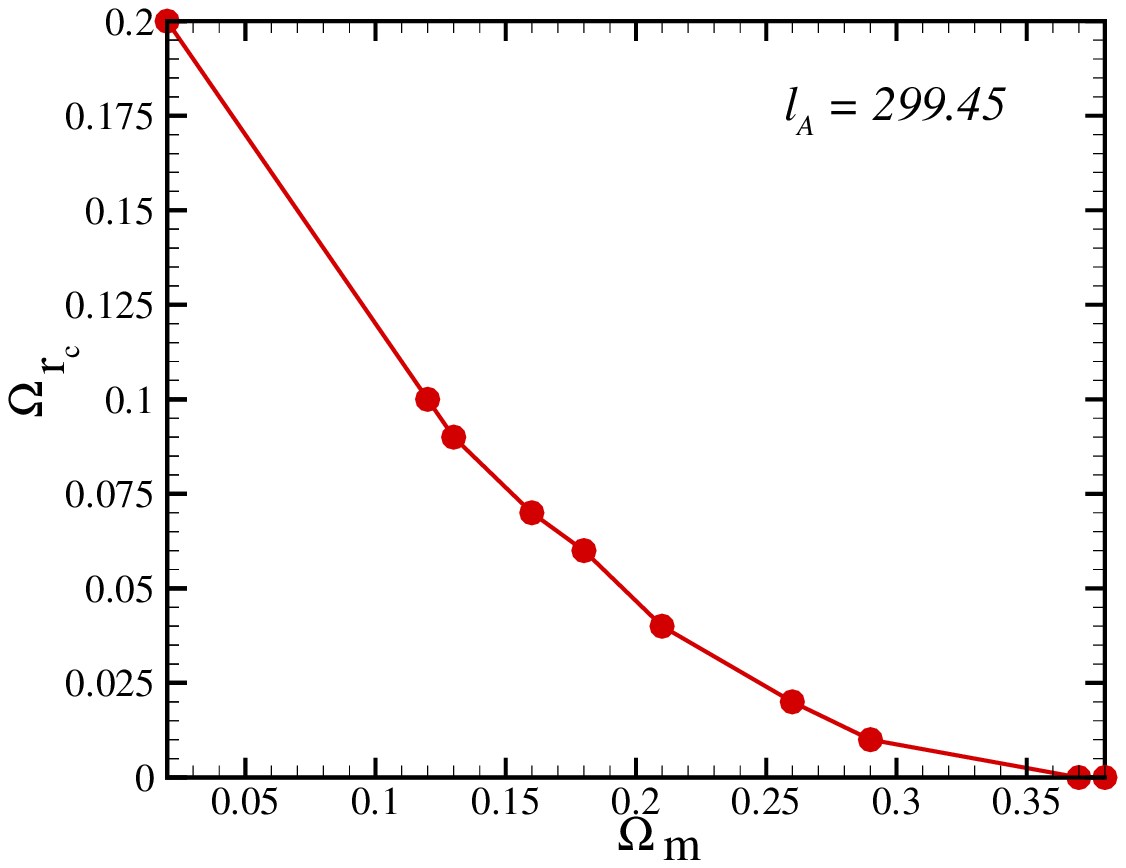}}
\centerline{\includegraphics[width=0.7\textwidth]{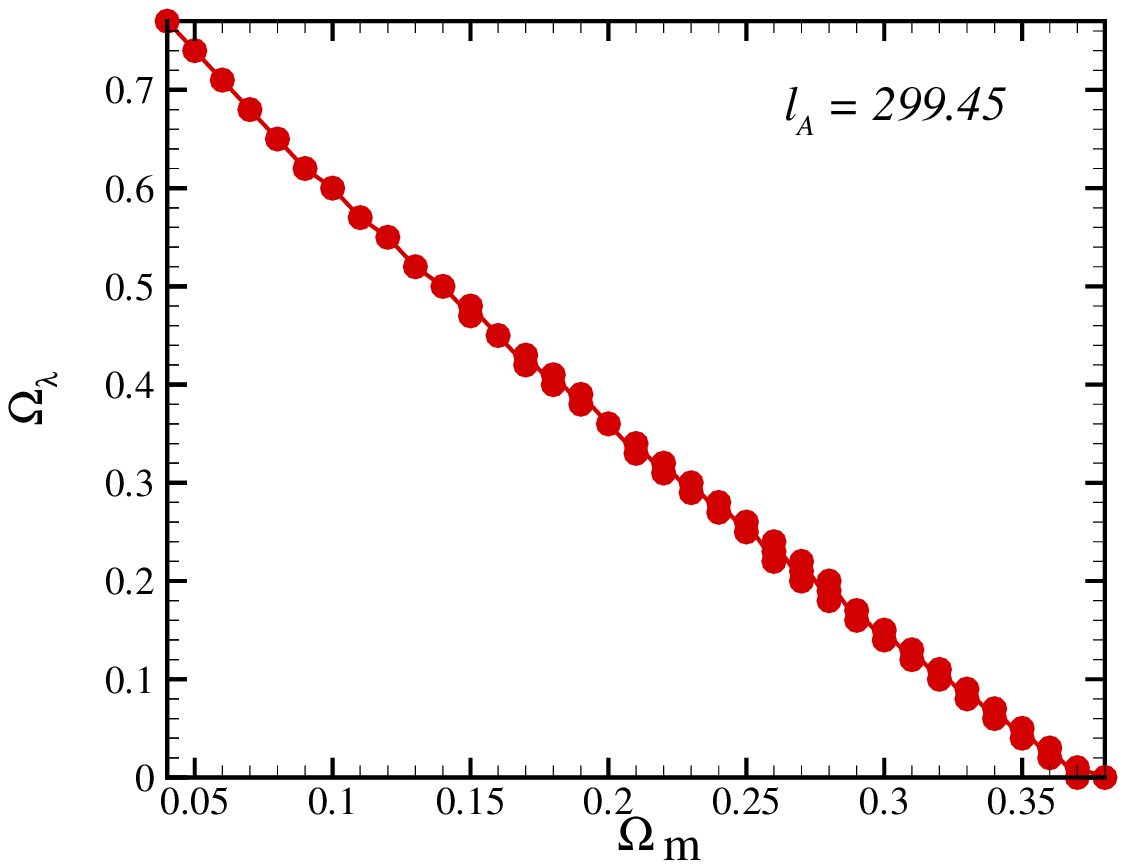}}
\caption{Constant acoustic angular scale in the joint space of
$\Omega_m$ and $\Omega_{r_c}$ (upper panel). Lower panel shows
dependence of acoustic angular scale on the $\Omega_m$ and
cosmological constant.} \label{l1}
\end{figure}
 Using $\chi^2_{\rm SNIa}+\chi_{l_1}^2$ analysis, we find the best fit values as:
$\Omega_m=0.15^{+0.03}_{-0.03}$,
$\Omega_{r_{c}}=0.08_{-0.02}^{+0.02}$ imply
$\Omega_K=0.42^{+0.10}_{-0.10}$ with $\chi^2_{min}/N_{d.o.f} =0.96$
for New catalog of Gold sample. For SNLS SNIa combined with the
position of first peak we find: $\Omega_m=0.11^{+0.01}_{-0.01}$,
$\Omega_{r_{c}}=0.11_{-0.01}^{+0.01}$ imply
$\Omega_K=0.36^{+0.05}_{-0.05}$ with $\chi^2_{min}/N_{d.o.f} =0.85$
at $1 \sigma$ level of confidence. Table \ref{peak} gives the best
values of parameters using the location of first and third peaks of
CMB power spectrum and other observational tests. We see that
$\Omega_m$ and $\Omega_{r_c}$ are very sensitive to the peaks
position of power spectrum of temperature fluctuations at the last
scattering surface. Since the phase transitions of peak position are
weakly model dependent we also apply the shift parameter of CMB to
extract best values for model parameters. According to the
$\chi_{\rm SNIa}^2+\chi_{\rm CMB}^2$ statistic we get:
$\Omega_m=0.23^{+0.04}_{-0.03}$,
$\Omega_{r_{c}}=0.14_{-0.01}^{+0.01}$ imply
$\Omega_K=0.03^{+0.08}_{-0.06}$ with $\chi^2_{min}/N_{d.o.f} =0.93$
for new catalog of Gold sample. For SNLS SNIa combined with the
position of first peak we find: $\Omega_m=0.17^{+0.02}_{-0.01}$,
$\Omega_{r_{c}}=0.16_{-0.01}^{+0.01}$ imply
$\Omega_K=0.05^{+0.05}_{-0.04}$ with $\chi^2_{min}/N_{d.o.f} =0.84$
at $1 \sigma$ level of confidence. The corresponding age of the
universe are $15.23_{-0.43}^{+0.50}$ and $14.48_{-0.19}^{+0.32}$,
respectively. These values are slightly different than that of
reported in the previous analysis \cite{zong,maartens1}.
\begin{table}[t]
\caption{The best values for the parameters of DGP self accelerating
model with the corresponding age for the universe from fitting with
SNIa from new Gold sample and SNLS data, SNIa+CMB+gas,
SNIa+CMB+gas+SDSS and SNIa+CMB+gas+SDSS+LSS experiments at one and
two $\sigma$ confidence level.} {\begin{tabular}{|c|c|c|c|c|} \hline
Observation & $\Omega_m$ & $\Omega_{r_{c}}$ &$\Omega_{K}$ & Age
(Gyr)
\\ \hline
  &&&& \\
 & $0.36^{+0.07}_{-0.06}$&$0.23^{+0.04}_{-0.04}$ & $-0.56^{+0.20}_{-0.20}$ & \\ 
 SNIa(new Gold)&&&&$13.78 ^{+0.68}_{-0.59}$\\
 & $0.36^{+0.13}_{-0.14}$&$0.23^{+0.09}_{-0.10}$
 &$-0.56^{+0.43}_{-0.48}$&
\\ &&&&\\ \hline
&&&&\\
SNIa(new Gold)& $0.23^{+0.04}_{-0.03}$&$0.14^{+0.02}_{-0.02}$ &$0.04^{+0.11}_{-0.10}$& \\
+CMB+gas &&&&$15.23^{+0.64}_{-0.45}$ \\
&$0.23^{+0.08}_{-0.06}$&$0.14^{+0.03}_{-0.03}$
&$0.04^{+0.18}_{-0.16}$&
 \\
 &&&&\\ \hline
&&&&\\
SNIa(new Gold)& $0.30^{+0.01}_{-0.05}$&$0.12^{+0.02}_{-0.01}$ &$0.01^{+0.09}_{-0.09}$& \\
+CMB+SDSS &&&&$14.52^{+0.15}_{-0.48}$ \\
+gas&$0.30^{+0.04}_{-0.06}$&$0.12^{+0.03}_{-0.03}$
&$0.01^{+0.15}_{-0.16}$&
 \\
 &&&&\\ \hline
 &&&& \\
SNIa(new Gold)& $0.28^{+0.03}_{-0.02}$&$0.13^{+0.01}_{-0.01}$
&$-0.002^{+0.064}_{-0.053}$&
 \\
 +CMB+SDSS&&&& $14.55^{+0.32}_{-0.22}$ \\
+LSS+gas&$0.28^{+0.05}_{-0.04}$&$0.13^{+0.02}_{-0.03}$ &$-0.002^{+0.117}_{-0.144}$& \\
 &&&&\\  
 \hline &&&&\\

  & $0.13^{+0.06}_{-0.06}$&$0.14^{+0.03}_{-0.03}$ & $0.20^{+0.16}_{-0.16}$ & \\ 
 SNIa (SNLS)&&&&$14.96^{+1.03}_{-1.43}$\\
 & $0.13^{+0.13}_{-0.12}$&$0.14^{+0.06}_{-0.07}$
 &$0.20^{+0.33}_{-0.35}$&
\\ &&&&\\  \hline
&&&&\\
SNIa(SNLS)& $0.17^{+0.01}_{-0.01}$&$0.16^{+0.01}_{-0.01}$ &$0.05^{+0.05}_{-0.05}$& \\
+CMB+gas &&&&$14.48^{+0.19}_{-0.19}$ \\
&$0.17^{+0.03}_{-0.04}$&$0.16^{+0.02}_{-0.02}$
&$0.05^{+0.10}_{-0.11}$&
 \\
 &&&&\\\hline
&&&&\\
SNIa(SNLS)& $0.22^{+0.01}_{-0.01}$&$0.15^{+0.01}_{-0.01}$ &$0.01^{+0.04}_{-0.04}$& \\
+CMB+SDSS &&&&$13.88^{+0.15}_{-0.15}$ \\
+gas&$0.22^{+0.03}_{-0.03}$&$0.15^{+0.02}_{-0.02}$
&$0.01^{+0.10}_{-0.10}$&
 \\
 &&&&\\ \hline
 &&&& \\
SNIa(SNLS)& $0.21^{+0.01}_{-0.01}$&$0.16^{+0.01}_{-0.01}$
&$0.01^{+0.04}_{-0.04}$&
 \\
 +CMB+SDSS&&&& $13.88^{+0.15}_{-0.15}$ \\
+LSS+gas&$0.21^{+0.03}_{-0.03}$&$0.16^{+0.02}_{-0.02}$
&$0.01^{+0.10}_{-0.10}$& \\
  \hline
\end{tabular}
\label{tab4} }
\end{table}

The recently detected size of baryonic peak in the SDSS is the third
observational data for our analysis.  The correlation function of
46,748 {\it Luminous Red Galaxies} (LRG) from the SDSS shows a well
detected baryonic peak around $100h^{-1}$ Mpc. This peak was
identified with the expanding spherical wave of baryonic
perturbations originating from acoustic oscillations at
recombination. This peak has an excellent match to the predicted
shape and the location of the imprint of the recombination-epoch
acoustic oscillation on the low-redshift clustering matter
\cite{eisenstein05}.  Recently Linder has shown in detail some
systematic uncertainties for baryon acoustic oscillation
\cite{linder}. Nonlinear mode coupling which is related to this fact
that the baryon acoustic oscillation is mostly contributed by linear
scale, but the influence of non-linear collapsing has quite broad
kernel. In other words, one might say that baryon acoustic
oscillation are $90-99\%$ linear in comparison to the CMB which is
$99.99\%$ linear, so this difference may affect on various models in
different way. Careful works to constrain on the free parameters of
underlying model needs to be carried out to determine the effect of
nonlinear mode coupling in the results of constraint by SDSS
observation. Nevertheless, roughly speaking regards the acceptance
intervals for free parameter cover the real intervals determined by
assuming nonlinearity mode for SDSS observation \cite{Amarzguioui}.

A dimensionless and independent of $H_0$ version of SDSS
observational parameter is:
\begin{eqnarray} \label{lss1}
{\cal A} &=&D_V(z_{\rm sdss})\frac{\sqrt{\Omega_mH_0^2}}{z_{\rm
sdss}}\nonumber\\
&=&\sqrt{\Omega_m}\left[\frac{H_0D_L^2(z_{\rm
sdss};\Omega_m,\Omega_{r_{c}})}{H(z_{\rm
sdss};\Omega_m,\Omega_{r_{c}})z_{\rm sdss}^2(1+z_{\rm
sdss})^2}\right]^{1/3}.
\end{eqnarray}
where $D_V(z_{\rm sdss})$ is characteristic distance scale of the
survey with mean redshift $z_{\rm sdss}$
\cite{eisenstein05,blak03,ness06}. We use the robust constraint on
the DGP model using the value of ${\cal A}=0.469\pm0.017$ from the
Luminous Red Galaxy (LRG) observation at $z_{\rm sdss} = 0.35$
\cite{eisenstein05}. This observation permits the addition of one
more term in the $\chi^2$ of equations (\ref{mar7}) and
(\ref{chi_cmb}) to be minimized with respect to $H(z)$ model
parameters. This term is:
\begin{equation}\label{chia}
\chi_{\rm SDSS}^2=\frac{\left[{\cal A}_{\rm obs}-{\cal A}_{\rm
th}\right]^2}{\sigma_{\rm sdss}^2}
\end{equation}

The baryon gas mass fraction for a range of redshifts is another
observational test, can also be used to constrain cosmological
models $H(z)$. The basic assumption corresponding to this method is
related to the baryon gas mass fraction in clusters
\cite{allen04,allen02,arn05} as:
\begin{equation}
S_{\rm gas}=\frac{M_{{\rm b}-{\rm gas}}}{M_{\rm tot}}
\end{equation}
this quantity is constant, related to the global fraction of the
universe $\Omega_b/\Omega_m$. $S_{\rm gas}$ can be written as:
\begin{equation}
S_{\rm gas}=\frac{1}{(1+\beta)}\frac{M_{{\rm b}}}{M_{\rm
tot}}=\frac{b}{(1+\beta)}\frac{\Omega_b}{\Omega_m}
\end{equation}
where $b$ is a bias factor suggesting that the baryon fraction in
clusters is slightly lower than for the universe as a whole. Also
$1+\beta$ is a factor taking into account the fact that the total
baryonic mass in clusters consists of both X-ray gas and optically
luminous baryonic mass (stars), the later being proportional to the
former with proportionality constant $\beta \simeq0.19\sqrt{h}$
\cite{allen04}. Dimensionless parameter for this observation is
given by \cite{ness06}:
\begin{eqnarray}
{\cal{S}}_{\rm
gas}(z;\Omega_m,\Omega_{r_c})&=&\frac{b}{1+\beta}\frac{\Omega_b}{\Omega_m}\left(\frac{D_A^{\rm
flat}(z)}{D_A(z;\Omega_m,\Omega_{r_c})}\right)^{\frac{3}{2}}\nonumber\\
&=&\xi\left(\frac{D_L^{\rm
flat}(z)}{D_L(z;\Omega_m,\Omega_{r_c})}\right)^{\frac{3}{2}}
\end{eqnarray}
where $D_A^{\rm flat}$ is the angular diameter distance
corresponding to flat pure CDM ($\Omega_m = 1$). Least square
quantity in the likelihood analysis for this observation is:
\begin{eqnarray}\label{chib1}
\chi^2(\xi,\Omega_m,\Omega_{r_{c}})&=&\sum_{i}\frac{[{\cal S}_{\rm gas}^{\rm obs}(z_i)-{\cal S}_{\rm gas}^{\rm th}(z_i;\Omega_m,\Omega_{r_{c}},\xi)]^2}{\sigma_i^2},\nonumber\\
\end{eqnarray}
Marginalizing over the nuisance parameter, $\xi$ gives:
\begin{eqnarray}\label{chib2}
\chi^2_{\rm gas}(\Omega_m,\Omega_{r_{c}})=K-\frac{W^2}{Y}
\end{eqnarray}
where
\begin{eqnarray}\label{chib3}
K=\sum_i\frac{{\cal S}_{\rm gas}^{\rm obs}(z_i)^2}{\sigma_i^2}
\end{eqnarray}

\begin{eqnarray}\label{chib4}
W=\sum_i\frac{{\cal S}_{\rm gas}^{\rm obs}(z_i)\cdot{\cal S}_{\rm
gas}^{\rm th}(z_i;\Omega_m,\Omega_{r_c},\xi=1)}{\sigma_i^2}
\end{eqnarray}
and
\begin{eqnarray}\label{chib5}
Y=\sum_i\frac{{\cal S}_{\rm gas}^{\rm
th}(z_i;\Omega_m,\Omega_{r_c},\xi=1)^2}{\sigma_i^2}
\end{eqnarray}
We use the $26$ cluster data for ${\cal{S}}_{\rm gas}^{\rm obs}(z)$
reported in Ref. \cite{allen04} to examine DGP modified gravity
model. According to  equations (\ref{mar7}), (\ref{chil1}),
(\ref{chil3}), (\ref{chi_cmb}), (\ref{chia}) and (\ref{chib2}) we
can constrain free parameters of the model using observational data
set related to background evolution.

In what follows we perform a combined analysis of SNIa, CMB, gas
cluster and SDSS to constrain the parameters of the DGP model by
minimizing the combined $\chi^2 = \chi^2_{\rm {SNIa}}+\chi^2_{{\rm
CMB}}+\chi^2_{\rm gas}+\chi^2_{{\rm SDSS}}$. The best values of the
model parameters from the fitting with the corresponding error bars
from the likelihood function marginalizing over the Hubble parameter
in the multidimensional parameter space are:
$\Omega_m=0.30_{-0.05}^{+0.01}$,
$\Omega_{r_{c}}=0.12_{-0.01}^{+0.02}$ and
$\Omega_K=0.01^{+0.09}_{-0.09}$ at $1\sigma$ confidence level with
$\chi^2_{min}/N_{d.o.f}=0.94$. The Hubble parameter corresponding to
the minimum value of $\chi^2$ is $h=0.61$. Here we obtain an age of
$14.52_{-0.48}^{+0.15}$ Gyr for the universe. Using the SNLS data,
the best fit values of model parameters are:
$\Omega_m=0.22_{-0.01}^{+0.01}$,
$\Omega_{r_{c}}=0.15_{-0.01}^{+0.01}$ and
$\Omega_K=0.01^{+0.04}_{-0.04}$ at $1\sigma$ confidence level with
$\chi^2_{min}/N_{d.o.f}=0.85$. Table \ref{tab4} indicates the best
fit values for the cosmological parameters with one and two $\sigma$
level of confidence.

Using the peaks position we find different values for the present
matter density and $\Omega_{r_c}$. Table \ref{peak} illustrates the
best fit values and corresponding derived age of universe. According
to the values reported in Tables \ref{tab4} and \ref{peak}, we infer
that the value of $\Omega_{r_c}$ is very sensitive to the
observational results from CMB. As usual we take the values confined
using shift parameter of CMB instead of one given by absolute values
from peaks position as a reliable results.

\begin{table}[t]
\caption{ The best values for the parameters of DGP self
accelerating model with the corresponding age for the universe from
fitting with SNIa from new Gold sample and SNLS data, SNIa+first
peak and SNIa+first peak+SDSS+LSS also for third peak experiments at
one and two $\sigma$ confidence level.}
{\begin{tabular}{|c|c|c|c|c|} \hline Observation & $\Omega_m$ &
$\Omega_{r_{c}}$ &$\Omega_{K}$ & Age (Gyr)
\\ \hline
&&&&\\
SNIa(new Gold)& $0.15^{+0.03}_{-0.03}$&$0.08^{+0.02}_{-0.02}$ &$0.42^{+0.10}_{-0.10}$& \\
+First Peak &&&&$15.86^{+0.48}_{-0.54}$ \\
&$0.15^{+0.06}_{-0.05}$&$0.08^{+0.03}_{-0.03}$
&$0.42^{+0.16}_{-0.16}$&
 \\
 &&&&\\ \hline
 &&&& \\
SNIa(new Gold)& $0.30^{+0.01}_{-0.01}$&$0.01^{+0.02}_{-0.04}$
&$0.57^{+0.08}_{-0.08}$&
 \\
 +First Peak+&&&& $13.77^{+0.09}_{-0.09}$ \\
SDSS+LSS&$0.30^{+0.02}_{-0.04}$&$0.01^{+0.02}_{-0.01}$ &$0.57^{+0.09}_{-0.09}$& \\
 &&&&\\  
\hline
&&&&\\
SNIa(new Gold)& $0.13^{+0.04}_{-0.01}$&$0.09^{+0.01}_{-0.02}$ &$0.41^{+0.08}_{-0.09}$& \\
+Third Peak &&&&$16.30^{+0.66}_{-0.28}$ \\
&$0.13^{+0.08}_{-0.02}$&$0.09^{+0.02}_{-0.05}$
&$0.41^{+0.16}_{-0.21}$&
 \\
 &&&&\\ \hline
 &&&& \\
SNIa(new Gold)& $0.29^{+0.01}_{-0.02}$&$0.01^{+0.01}_{-0.01}$
&$0.58^{+0.08}_{-0.08}$&
 \\
 +Third Peak+&&&& $13.85^{+0.09}_{-0.09}$ \\
SDSS+LSS&$0.29^{+0.02}_{-0.07}$&$0.01^{+0.02}_{-0.01}$ &$0.58^{+0.15}_{-0.11}$& \\
 &&&&\\  
\hline
&&&&\\
SNIa(SNLS)& $0.11^{+0.01}_{-0.01}$&$0.11^{+0.01}_{-0.01}$ &$0.36^{+0.05}_{-0.05}$& \\
+First Peak &&&&$14.99^{+0.22}_{-0.23}$ \\
&$0.11^{+0.02}_{-0.02}$&$0.11^{+0.02}_{-0.02}$
&$0.36^{+0.09}_{-0.09}$&
 \\
 &&&&\\ \hline
 &&&&\\ 
SNIa(SNLS)& $0.15^{+0.01}_{-0.01}$&$0.08^{+0.01}_{-0.01}$
&$0.42^{+0.05}_{-0.05}$&
 \\
 +First Peak+&&&& $14.25^{+0.16}_{-0.16}$ \\
SDSS+LSS&$0.15^{+0.02}_{-0.03}$&$0.08^{+0.02}_{-0.02}$ &$0.42^{+0.09}_{-0.09}$& \\
 &&&&\\  
  \hline
&&&&\\
SNIa(SNLS)& $0.08^{+0.01}_{-0.01}$&$0.13^{+0.01}_{-0.01}$ &$0.33^{+0.04}_{-0.04}$& \\
+Third Peak &&&&$15.91^{+0.29}_{-0.31}$ \\
&$0.08^{+0.02}_{-0.02}$&$0.13^{+0.02}_{-0.02}$
&$0.33^{+0.09}_{-0.09}$&
 \\
 &&&&\\ \hline
 &&&& \\
SNIa(SNLS)& $0.16^{+0.03}_{-0.03}$&$0.07^{+0.02}_{-0.01}$
&$0.45^{+0.10}_{-0.09}$&
 \\
 +Third Peak+&&&& $14.04^{+0.40}_{-0.43}$ \\
SDSS+LSS&$0.16^{+0.04}_{-0.04}$&$0.07^{+0.03}_{-0.02}$ &$0.45^{+0.14}_{-0.11}$& \\
 &&&&\\
 \hline
 \end{tabular}
\label{peak}}
\end{table}
\begin{figure}[t]
\begin{center}
\centerline{\includegraphics[width=0.7\textwidth]{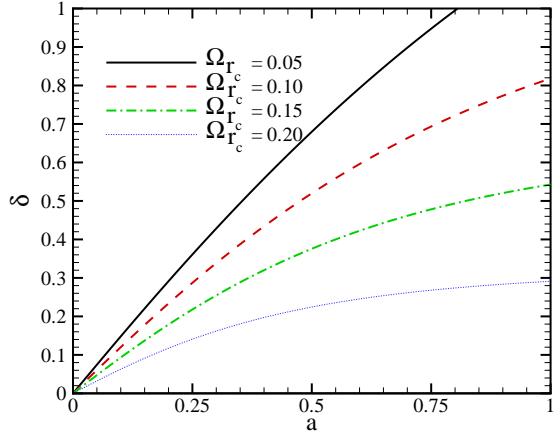}}
\caption{Evolution of density contrast as a function of scale factor
for different values of $\Omega_{r_{c}}$ in a flat universe.}
\label{ev}
\end{center}
\end{figure}
\begin{figure}[t]
\begin{center}
\centerline{\includegraphics[width=0.7\textwidth]{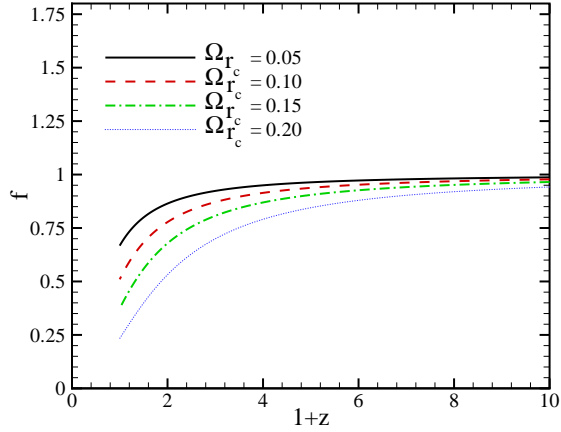}}
\caption{Growth index versus redshift for different values of
$\Omega_{r_{c}}$. Here we imagined   flat universe.} \label{4}
\end{center}
\end{figure}
\section{Constraints by Large Scale structure }
\label{cstructure}
So far we have only considered observational
results related to the background evolution. In this section using
the linear approximation of structure formation we obtain the
growth index of structures and compare it with the result of
observations by the $2$-degree Field Galaxy Redshift Survey
($2$dFGRS).

Koyama and Maartens \cite{maartens2} have recently shown the
evolution of density perturbations requires an analysis of the
5-dimensional gravitational field. In this model Poisson equation is
modified and shows the suppression of growth due to gravity leakage.
The continuity and modified Poisson equations for the density
contrast $\delta=\delta \rho/\bar{\rho}$ in the cosmic fluid provide
the evolution of density contrast in the linear approximation (i.e.
$\delta\ll 1 $)\cite{maartens2,P93,B04} as:
\begin{equation}
\ddot{\delta}+2\frac{\dot{a}}{a}\dot{\delta} - \left[ {v_s}^2
\nabla^2 +4\pi G \left(1 + \frac{1}{3 \alpha} \right)\rho\right]
\delta=0, \label{eq2}
\end{equation}
 where
\begin{equation}
\alpha =1 -2 r_c H \left(1+ \frac{\dot{H}}{3 H^2} \right)
\end{equation}
 the dot denotes the derivative with respect to time.
 Thus the growth rate receives an additional modification from the
time variation of Newton's constant through $\alpha$.

The effect of dark energy in the evolution of the structures in this
equation enters through its influence on the expansion rate.
 The validity of this linear Newtonian approach is
restricted to perturbations on the sub-horizon scales but large
enough where structure formation is still in the linear regime
\cite{maartens2,P93,B04}. For the perturbations larger than the
Jeans length, $ \lambda_J= \pi^{1/2} v_s / \sqrt{G\left(1 +
\frac{1}{3 \alpha} \right) \rho} $, equation (\ref{eq2}) for cold
dark matter (CDM) reduces to:
\begin{equation} \ddot{\delta}+2 \frac{\dot{a}}{a} \dot{\delta}-4\pi G\left(1 +
\frac{1}{3 \alpha} \right) \rho\delta=0 \label{eq3}
\end{equation}
The equation for the evolution of density contrast can be
rewritten in terms of the scale factor as:
\begin{equation}
\frac{d^2\delta}{da^2}+\frac{d\delta}{da}\left[\frac{\ddot{a}}{\dot{a}^2}+\frac{2H}{\dot{a}}\right]-
\frac{3H_0^2}{2\dot{a}^2a^3}\left(1 + \frac{1}{3 \alpha}
\right)\Omega_{m}\delta=0. \label{eq31}
\end{equation}
Numerical solution of equation (\ref{eq31}) in the FRW universe in
the background of DGP model is shown in Figure \ref{ev}. In the CDM
model, the density contrast $\delta$ grows linearly with the scale
factor, while we have a deviation from the linearity as soon as
universe enters to acceleration era. Increasing $\Omega_{r_c}$ leads
a decreasing in the evolution of density contrast which is in
agreement to the finding about the behavior acceleration parameter
versus $\Omega_{r_c}$ (see Figure \ref{acc}).

In the linear perturbation theory, the peculiar velocity field
$\bf{v}$ is determined by the density contrast \cite{P93,P80} as:
\begin{equation} {\bf v} ({\bf x})= H_0 \frac{f}{4\pi} \int \delta ({\bf y})
\frac{{\bf x}-{\bf y}}{\left| {\bf x}-{\bf y} \right|^3} d^3 {\bf
y}, \end{equation} where the growth index $f$ is defined by:
\begin{equation} f=\frac{d \ln \delta}{d \ln a},
 \label{eq5}
\end{equation}
and it is proportional to the ratio of the second term of equation
(\ref{eq3}) (friction) to the third (Poisson) term.

We use the evolution of the density contrast $\delta$ to compute the
growth index of structure $f$, which is an important quantity for
the interpretation of peculiar velocities of galaxies, as discussed
in \cite{P80,rah02} for the Newtonian and the relativistic regime of
structure formation. Replacing the density contrast with the growth
index in equation (\ref{eq31}) results in the evolution of growth
index as:
\begin{eqnarray}
\label{index} &&\frac{df}{d\ln a}= \frac{3H_0^2}{2\dot{a}^2a}\left(1
+ \frac{1}{3 \alpha} \right)\Omega_m
-f^2-f\left[1+\frac{\ddot{a}}{aH^2}\right]
\end{eqnarray}
Figure \ref{4} shows the numerical solution of equation
(\ref{index}) in terms of redshift. An analytic formula for the
present growth index in the flat $\Lambda$CDM model has been given
in ref. \cite{f} as $f(z=0.0,\Omega_m)\simeq \Omega_m^{0.6}$, here
we extend this formula for the universe governed by DGP modified
gravity. The simplest form for fitting formula in the wide range of
$\Omega_m$ and $\Omega_{r_c}$ is
\begin{eqnarray}\label{fit}
f(z=0.0;\Omega_m,\Omega_{r_c})&\simeq&
\Omega_m^{0.6}+\left(0.0159+0.0603\Omega_m\right)\times\nonumber\\
&&\exp\biggl(\left[1.0694-0.3867\ln\Omega_m\right]^2\Omega_{r_c}\biggr)\nonumber\\
&&-0.0542\Omega_m-0.0100
\end{eqnarray}
Figure \ref{fit1} shows growth index, $f(z=0.0;\Omega_m=0.30)$, as a
function of $\Omega_{r_c}$ derived from numerical solution of
equation (\ref{index}) and illustrated by fitting formula (equation
(\ref{fit})).

\begin{figure}[t]
\begin{center}
\centerline{\includegraphics[width=0.7\textwidth]{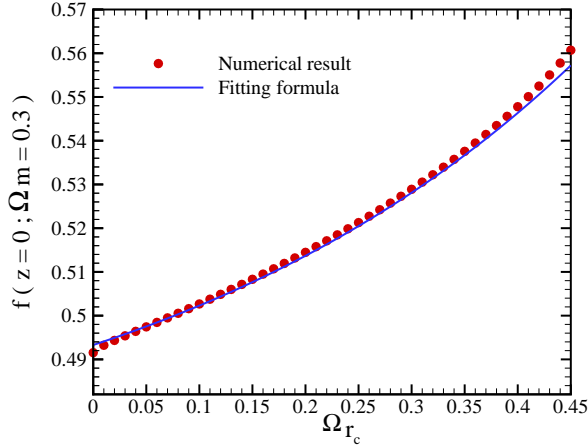}}
\caption{Growth index versus $\Omega_{r_c}$ for $z=0.0$ and
$\Omega_m=0.30$. Solid line is given by fitting formula and symbol
is derived by numerical solution of equation (\ref{index}). }
\label{fit1}
\end{center}
\end{figure}

To use observational results implied to linear structure formation
we rely to the observation of $220,000$ galaxies with the $2$dFGRS
experiment provides the numerical value of growth index
\cite{eisenstein05}. By measurements of two-point correlation
function, the $2$dFGRS team reported the redshift distortion
parameter of $\varepsilon = f/\kappa =0.49\pm0.09$ at $z=0.15$,
where $\kappa$ is the bias parameter describing the difference in
the distribution of galaxies and their masses. Verde et al. (2003)
used the bispectrum of $2$dFGRS galaxies \cite{ver01,la02} and
obtained $\kappa_{verde} =1.04\pm 0.11$ which gave $f= 0.51\pm0.10$.
Now we fit the growth index at the present time derived from the
equation (\ref{index}) with the observational value. This fitting
gives a less constraint to the parameters of the model, so in order
to have a better confinement of the parameters, we combine this
fitting with those of SNIa$+$CMB$+$SDSS which have been discussed in
the previous section. We perform the least square fitting by
minimizing $\chi^2=\chi_{\rm{SNIa}}^2+\chi_{\rm{CMB}}^2+\chi_{\rm
gas}^2+\chi_{\rm{SDSS}}^2+\chi_{\rm{LSS}}^2$, where
\begin{eqnarray}
\chi^2_{\rm
{LSS}}=\frac{[f_{obs}(z=0.15)-f_{th}(z=0.15;\Omega_m,\Omega_{r_{c}})]^2}{\sigma_{f_{obs}}^2}
\end{eqnarray}
 The best fit values with the corresponding error bars for the
model parameters by using new Gold sample data are:
$\Omega_m=0.28^{+0.03}_{-0.02}$,
$\Omega_{r_{c}}=0.13_{-0.01}^{+0.01}$ and
$\Omega_K=-0.002_{-0.053}^{+0.064}$ at $1\sigma$ confidence level
with $\chi^2_{min}/N_{d.o.f}=0.93$. Using the SNLS supernova data,
the best fit values for model parameters are:
$\Omega_m=0.21^{+0.01}_{-0.01}$,
$\Omega_{r_{c}}=0.16_{-0.01}^{+0.01}$ and $\Omega_K=0.01_{-0.04
}^{+0.04}$ at $1\sigma$ confidence level with
$\chi^2_{min}/N_{d.o.f}=0.84$. The error bars have been obtained
through the likelihood functions $(\mathcal{L}$$\propto
e^{-\chi^2/2})$ marginalized over the nuisance parameter $h$
\cite{press94}. 
The best values reported in the Ref. \cite{maartens1} using
SNIa+CMB+SDSS are: $\Omega_m=0.270$, $\Omega_{r_{c}}=0.125$ for Gold
sample SNIa and for SNLS SNIa are: $\Omega_m=0.255$,
$\Omega_{r_{c}}=0.130$, while in Ref. \cite{zong} using SNIa+SDSS,
$\Omega_m=0.270_{+0.018}^{-0.017}$,
$\Omega_{r_{c}}=0.216_{-0.013}^{+0.012}$. We concluded that
observational results from large scale structure given by $2$dfGRS
in addition to including baryon gas mass fraction results, put weak
constraints on the DGP model free parameters.

The likelihood functions for the three cases of (i) fitting model
with Supernova data, (ii) combined analysis with the three
experiments of SNIa$+$CMB$+$gas and (iii) combining all five
experiments of SNIa$+$CMB$+$gas$+$SDSS$+$LSS are shown in Figures
\ref{mbhw} and \ref{mbhw1}. The joint confidence contours in the
$(\Omega_m,\Omega_{r_{c}})$ plane are also shown in Figures
\ref{jmw1} and \ref{jmw11} for Gold sample and combined
observational results, respectively. Figures \ref{jmw2} and
\ref{jmw22} show the joint confidence interval for SNLS data and
SNIa$+$CMB$+$gas$+$SDSS$+$LSS experiments.

\section{Age of Universe}\label{agesection} The "age crisis" is one the main reasons
of the acceleration phase of the universe. The problem is that the
universe's age in the Cold Dark Matter (CDM) universe is less than
the age of old stars in it. Studies on the old stars
\cite{carretta00} suggest an age of $13^{+4}_{-2}$ Gyr for the
universe. Richer et. al. \cite{richer02} and Hasen et. al.
\cite{hansen02} also proposed an age of $12.7\pm0.7$ Gyr, using the
white dwarf cooling sequence method (for full review of the cosmic
age see \cite{spe03}). The age of universe integrated from the big
bang up to now is given by:
\begin{eqnarray}\label{age}
t_0(\Omega_m,\Omega_{r_{c}}) &=& \int_0^{t_0}\,dt\nonumber \\
&=&\int_0^\infty {dz'\over (1+z') H(z')}
\end{eqnarray}
Figure~\ref{fig:1} shows the dependence of $H_0t_0$ (Hubble
parameter times the age of universe) on $\Omega_{r_{c}}$ for a flat
universe. Obviously increasing $\Omega_{r_{c}}$ results in a longer
age for the universe. As shown in the lower panel of Figure
\ref{fig:1}, $\Omega_{r_c}$ behaves as the same as dark energy,
$\Omega_{\Lambda}$, in the flat $\Lambda$CDM model.
\begin{figure}[t]
\centerline{\includegraphics[width=0.7\textwidth]{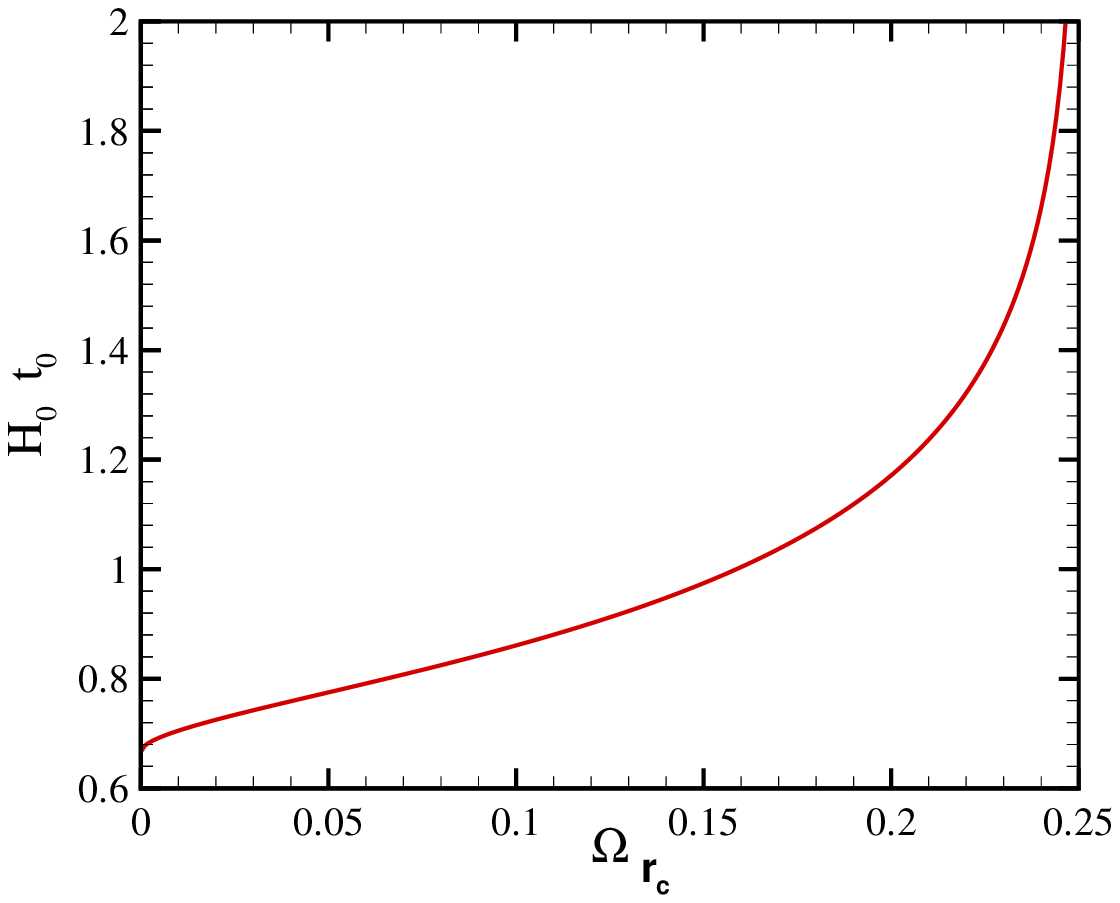}}
\centerline{\includegraphics[width=0.7\textwidth]{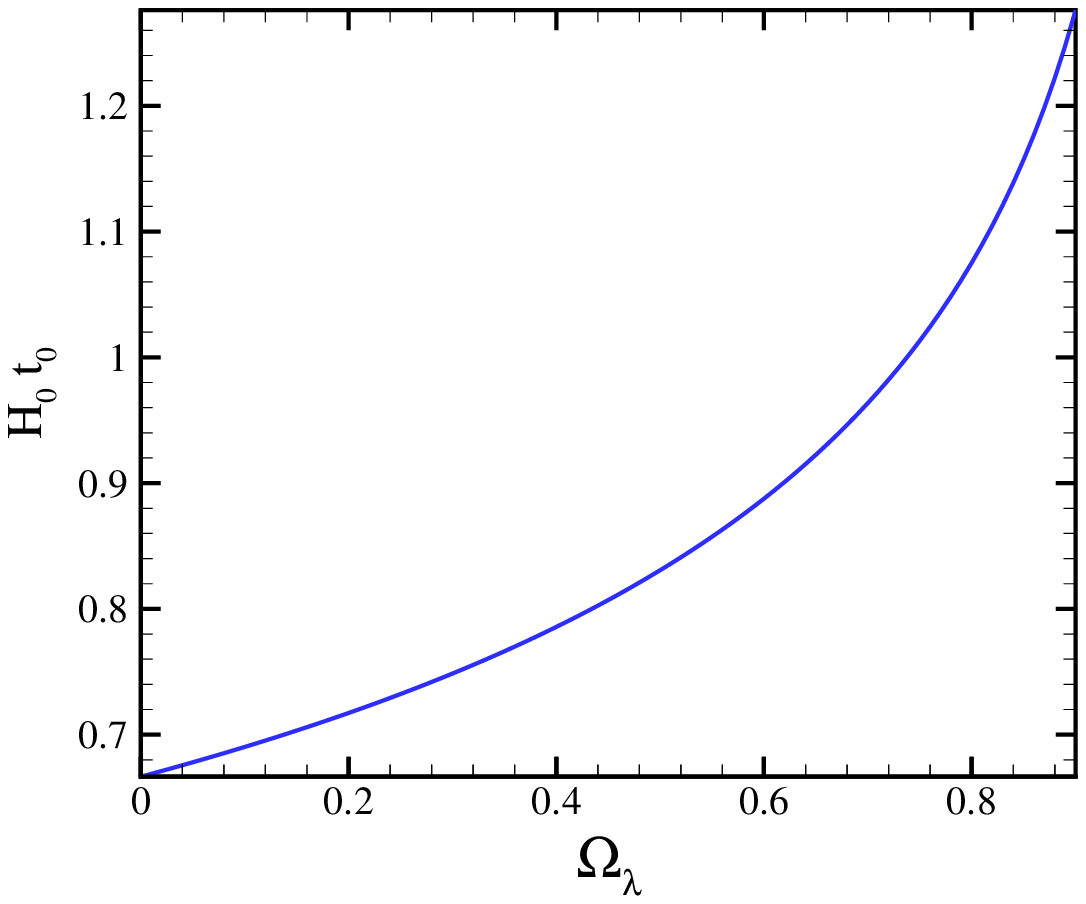}}
\caption{ $H_0t_0$ (age of universe times the Hubble constant at the
present time) as a function of $\Omega_{r_{c}}$ in a flat universe.
Increasing $\Omega_{r_{c}}$ gives a longer age for the universe.
Lower panel shows the same function versus $\Omega_{\lambda}$ in the
flat $\Lambda$CDM model and $w=-1.0$. } \label{fig:1}
 \end{figure}
Finally we do the consistency test, comparing the age of universe
derived from this model with the age of old stars and Old High
Redshift Galaxies (OHRG) in various redshifts.  Table \ref{tab4}
shows that the age of universe from the combined analysis of
SNIa$+$CMB$+$gas$+$SDSS$+$LSS is $14.55_{-0.22}^{+0.32}$ Gyr and
$13.88_{-0.15}^{+0.15}$ for new Gold sample and SNLS data,
respectively. These values are in agreement with the age of old
stars \cite{carretta00}. Here we take three OHRG for comparison with
the DGP model, namely the LBDS $53$W$091$, a $3.5$-Gyr old radio
galaxy at $z=1.55$ \cite{dunlop96}, the LBDS $53$W$069$ a $4.0$-Gyr
old radio galaxy at $z=1.43$ \cite{dunlop99} and a quasar, APM
$08279+5255$ at $z=3.91$ with an age of $t=2.1_{-0.1}^{+0.9}$Gyr
\cite{hasinger02}. The latter has once again led to the "age
crisis". An interesting point about this quasar is that it cannot be
accommodated in the $\Lambda$CDM model \cite{jan06}. To quantify the
age-consistency test we introduce the expression $\tau$ as:
\begin{equation}
 \tau=\frac{t(z;\Omega_m,\Omega_{r_{c}})}{t_{obs}} = \frac{t(z;\Omega_m,\Omega_{r_{c}})H_0}{t_{obs}H_0},
\end{equation}
where $t(z)$ is the age of universe, obtained from the equation
(\ref{age}) and $t_{obs}$ is an estimation for the age of old
cosmological object. In order to have a compatible age for the
universe we should have $\tau>1$. Table \ref{tab6} shows the value
of $\tau$ for three mentioned OHRG. We see that the parameters of
DGP model from the combined observations don't provide a compatible
age for the universe, compared to the age of old objects, while  the
SNLS data result in a longer age for the universe. Once again for
the DGP model, APM $08279+5255$ at $z=3.91$ has a longer age than
the universe but gives better results than most Quintessence and
braneworld models \cite{sa1,sa2,sa3,sa4}.

\begin{table}[t]
\caption{The value of $\tau$ for three high redshift objects, using
the parameters of the model derived from fitting with the
observations.} {\begin{tabular}{|c|c|c|c|}
 \hline
  Observation & LBDS &LBDS  & APM  \\
& $53$W$069$&$53$W$091$& $08279+5255$ \\
  & $z=1.43$&$z=1.55$& $z=3.91$  \\ \hline
  &&&\\
SNIa (new Gold)& $1.37^{+0.12}_{-0.12}$ & $1.12^{+0.10}_{-0.10}$& $0.79^{+0.07}_{-0.07}$ \\
&&&\\
 \hline

&&&\\
SNIa(new Gold)+CMB &$1.35^{+0.03}_{-0.08}$&$1.11^{+0.03}_{-0.07}$&$0.83^{+0.02}_{-0.06}$ \\
 +SDSS+gas& && \\
\hline
&&&\\
SNIa(new Gold)+CMB & $1.37^{+0.05}_{-0.03}$&$1.12^{+0.04}_{-0.03}$&$0.84^{+0.03}_{-0.03}$ \\
 +SDSS+LSS+gas& && \\ \hline
&&&\\
 SNIa (SNLS)& $ 1.53^{+0.17}_{-0.23}$ & $1.26^{+0.14}_{-0.19}$& $1.00^{+0.13}_{-0.21}$ \\ &&&\\ \hline
&&&\\
SNIa(SNLS)+CMB & $1.36^{+0.03}_{-0.03}$&$1.12^{+0.02}_{-0.02}$&$0.85^{+0.02}_{-0.02}$ \\
 +SDSS+gas& && \\
\hline
&&&\\
SNIa(SNLS)+CMB & $1.36^{+0.03}_{-0.03}$&$1.12^{+0.02}_{-0.02}$&$0.85^{+0.02}_{-0.02}$ \\
 +SDSS+LSS+gas& && \\
\hline
\end{tabular}
\label{tab6} }
\end{table}

\begin{figure}[t]
\centerline{\includegraphics[width=0.7\textwidth]{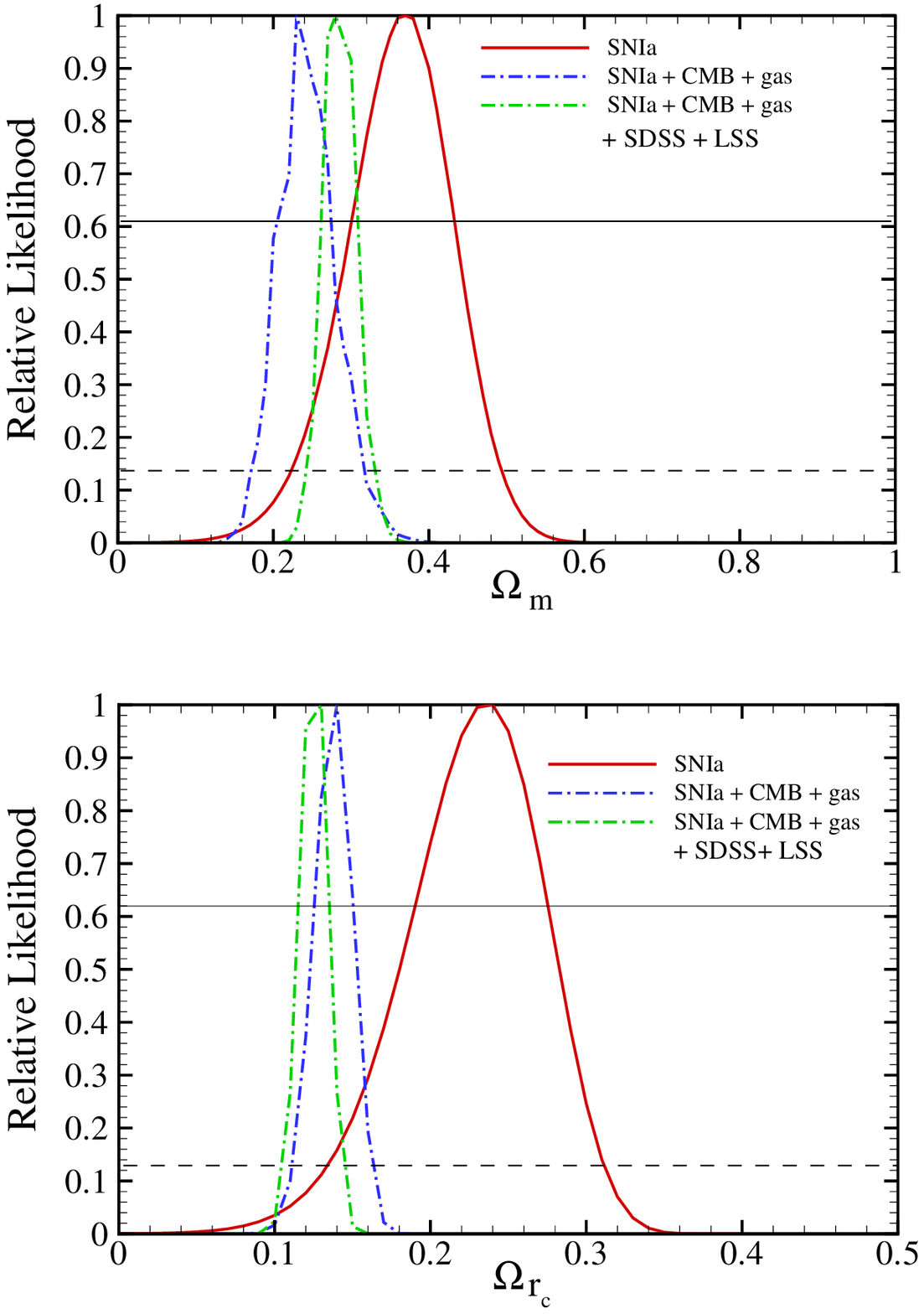}}
\caption{Marginalized likelihood functions of two parameters of DGP
model ($\Omega_m$ and $\Omega_{r_{c}}$). The solid line corresponds
to the likelihood function of fitting the model with SNIa data (new
Gold sample), the dashdot line with the joint SNIa$+$CMB$+$gas data
and dashed line corresponds to SNIa$+$CMB$+$gas$+$SDSS$+$LSS. The
intersections of the curves with the horizontal solid and dashed
lines give the bounds with $1\sigma$ and $2\sigma$ level of
confidence respectively.} \label{mbhw}
 \end{figure}


\begin{figure}[t]
\centerline{\includegraphics[width=0.7\textwidth]{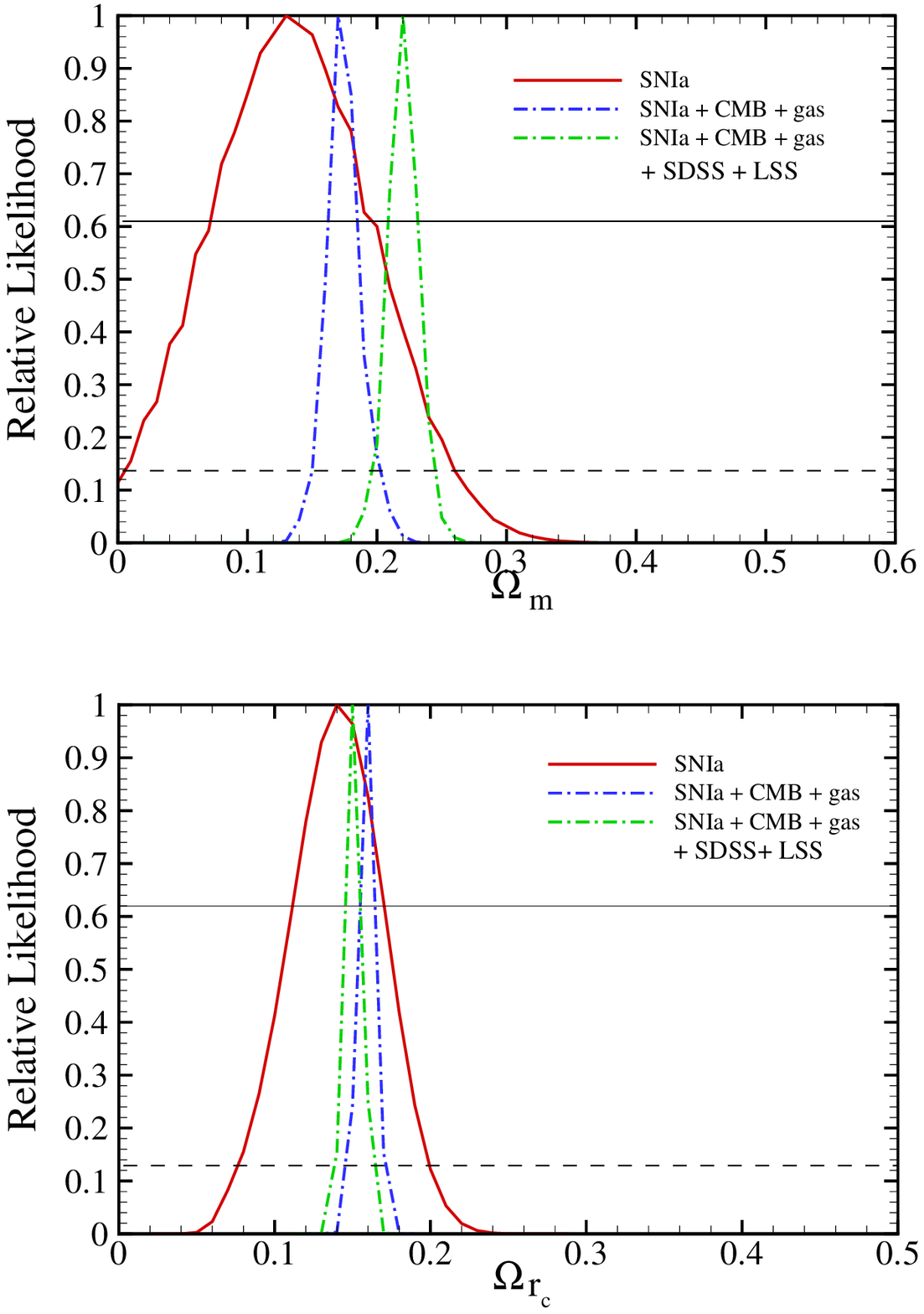}}
\caption{Marginalized likelihood functions of two parameters of DGP
model ($\Omega_m$ and $\Omega_{r_{c}}$). The solid line corresponds
to the likelihood function of fitting the model with SNIa data
(SNLS), the dashdot line with the joint SNIa$+$CMB$+$gas data and
dashed line corresponds to SNIa$+$CMB$+$gas$+$SDSS$+$LSS. The
intersections of the curves with the horizontal solid and dashed
lines give the bounds with $1\sigma$ and $2\sigma$ level of
confidence respectively.} \label{mbhw1}
 \end{figure}

\begin{figure}[t]
\centerline{\includegraphics[width=0.7\textwidth]{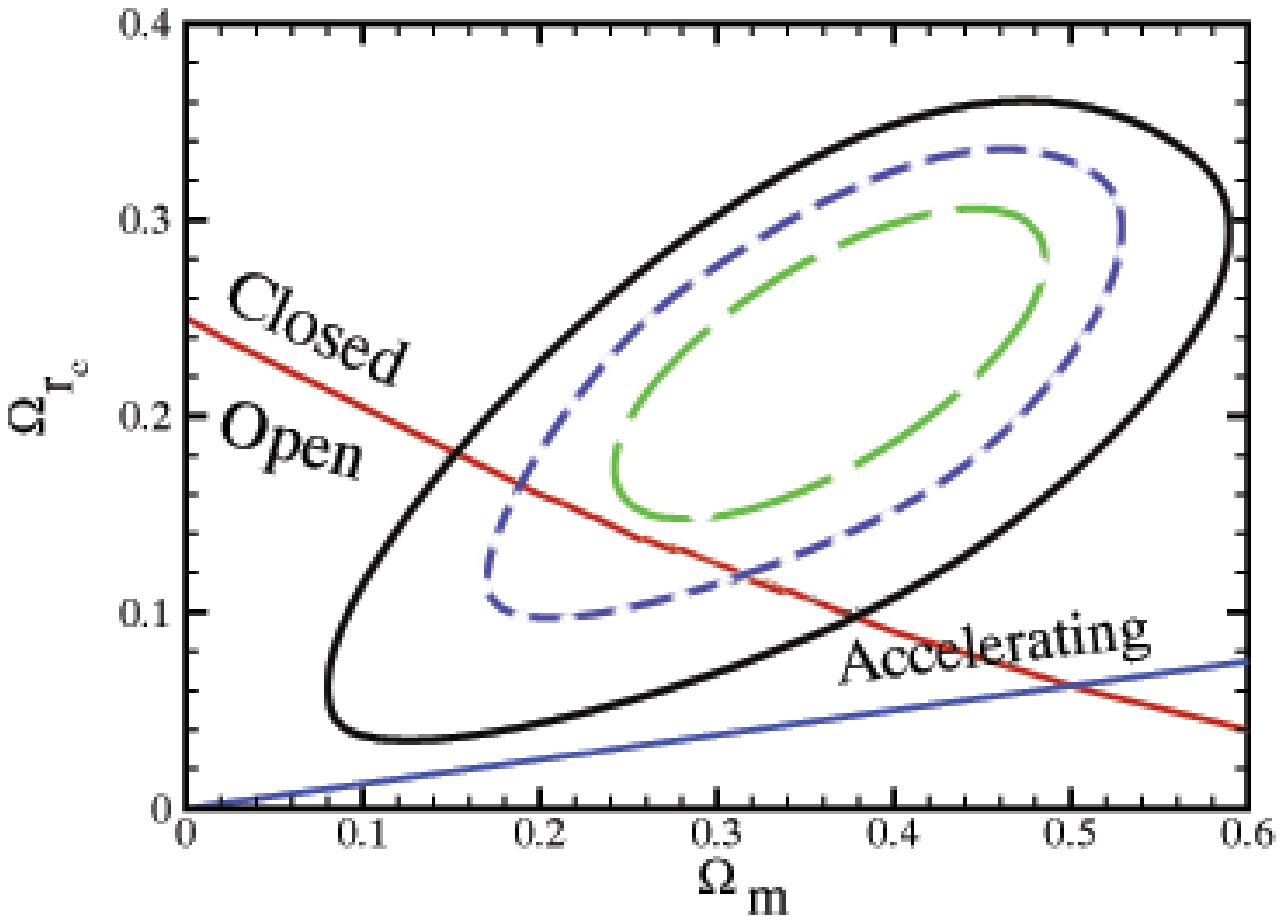}}
\caption{Joint confidence intervals of $\Omega_m$ and
$\Omega_{r_{c}}$, fitted with SNIa new Gold sample. Solid line,
dashed line and long dashed line correspond to $3\sigma$, $2\sigma$
and $1\sigma$ level of confidence, respectively.} \label{jmw1}
 \end{figure}

 \begin{figure}[t]
\centerline{\includegraphics[width=0.7\textwidth]{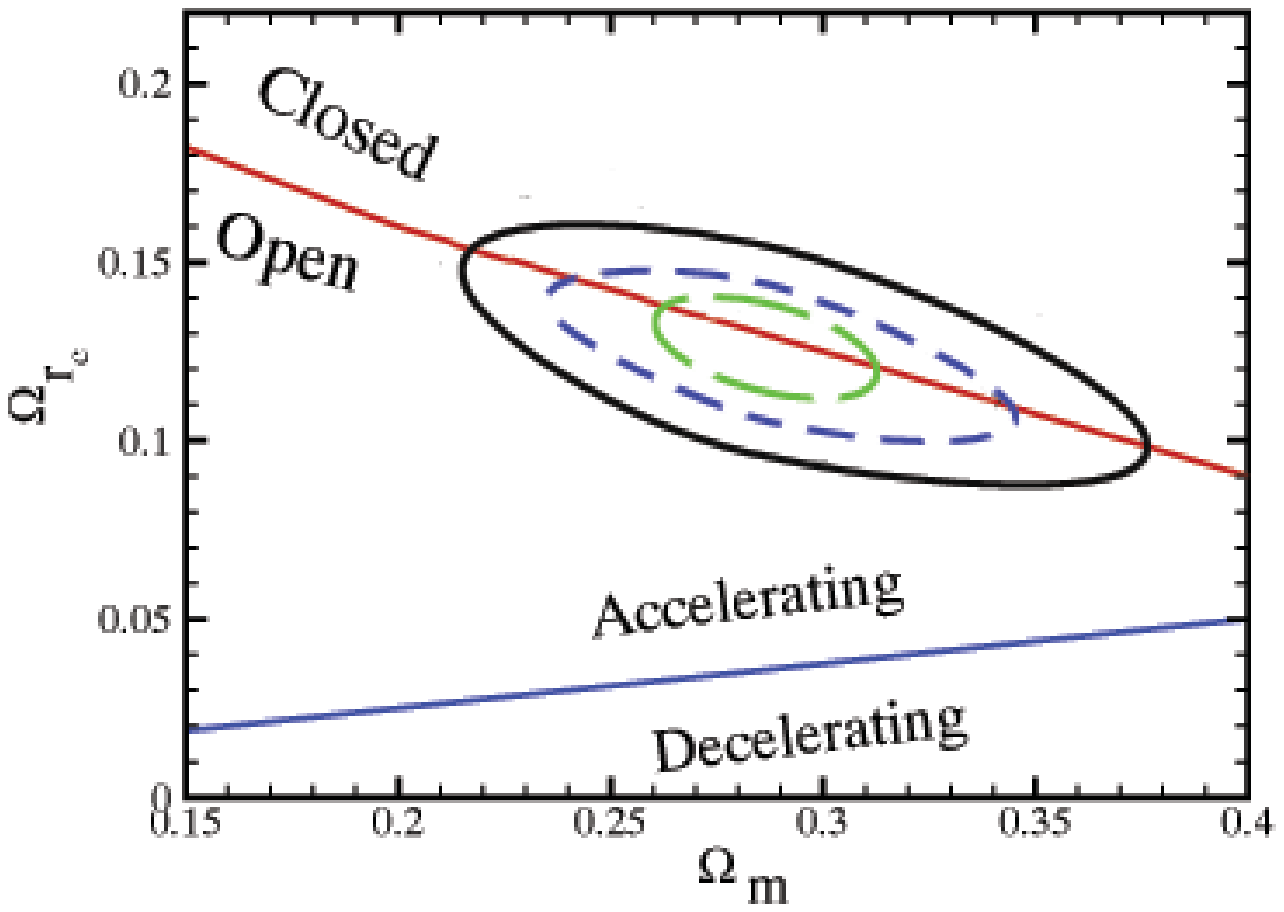}}
\caption{Joint confidence intervals of $\Omega_m$ and
$\Omega_{r_{c}}$, fitted with SNIa new Gold
sample$+$CMB$+$gas$+$SDSS$+$LSS. Solid line, dashed line and long
dashed line correspond to $3\sigma$, $2\sigma$ and $1\sigma$ level
of confidence, respectively.} \label{jmw11}
 \end{figure}
\begin{figure}
\centerline{\includegraphics[width=0.7\textwidth]{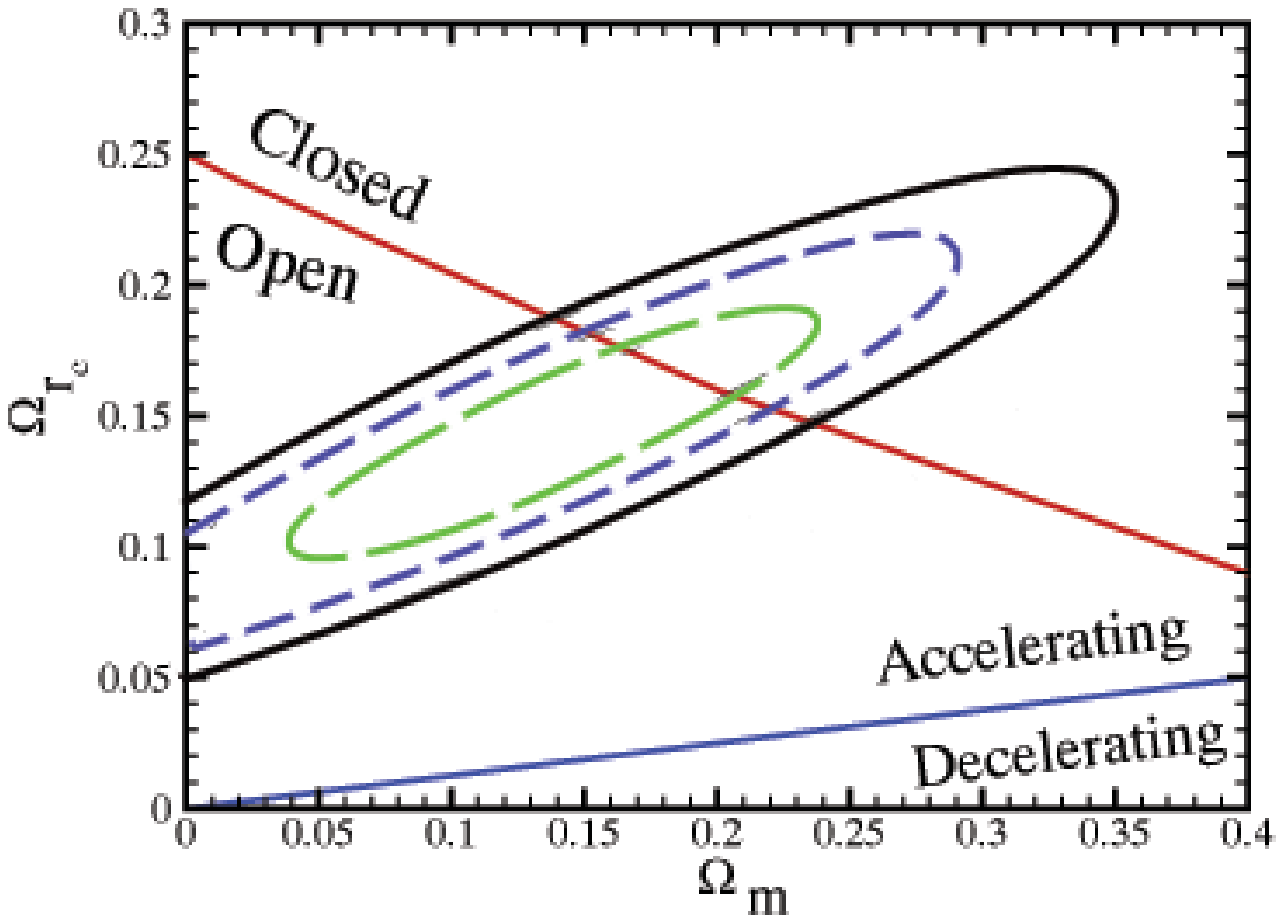}}
\caption{Joint confidence intervals of $\Omega_m$ and
$\Omega_{r_{c}}$, fitted with SNIa SNLS. Solid line, dashed line and
long dashed line correspond to $3\sigma$, $2\sigma$ and $1\sigma$
level of confidence, respectively.} \label{jmw2}
 \end{figure}

 \begin{figure}
\centerline{\includegraphics[width=0.7\textwidth]{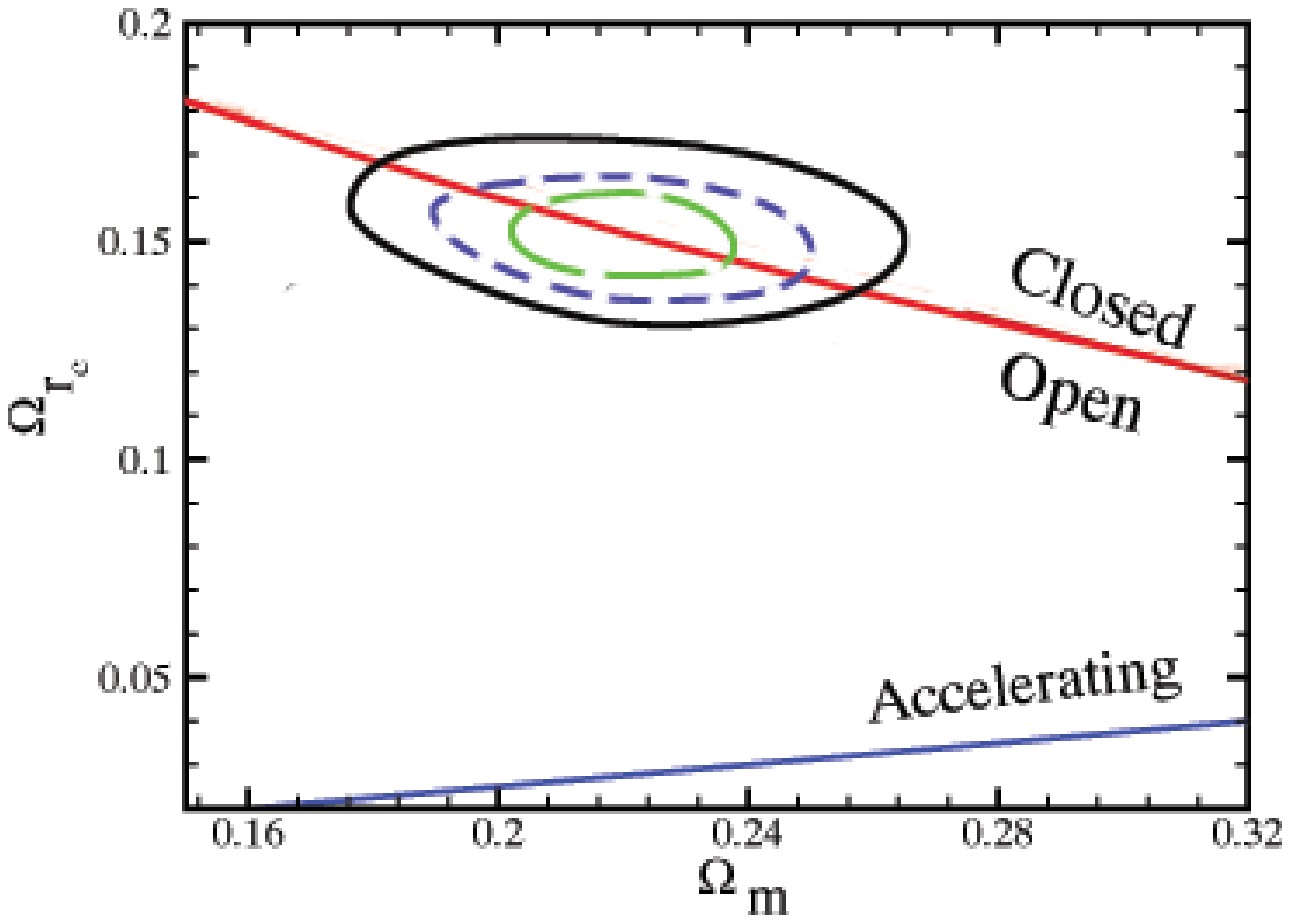}}
\caption{Joint confidence intervals of $\Omega_m$ and
$\Omega_{r_{c}}$, fitted with SNIa SNLS$+$CMB$+$gas$+$SDSS$+$LSS.
Solid line, dashed line and long dashed line correspond to
$3\sigma$, $2\sigma$ and $1\sigma$ level of confidence,
respectively.} \label{jmw22}
 \end{figure}

\section{Conclusion}
\label{conc} We studied a self accelerating cosmological model, DGP
modified gravity. The effect of this model on the age of universe,
the radial comoving distance, the comoving volume element and the
variation of the apparent size of objects with the redshift
(Alcock-Paczynski test) have been studied. The evolution of density
contrast, $\delta$, as a function of scale factor for various values
of $\Omega_{r_c}$ shows that increasing $\Omega_{r_c}$ suppresses
the growth of density contrast, which is in agreement with the
behavior of acceleration parameter versus $\Omega_{r_c}$. We
extrapolate the relation of the growth factor in terms of $\Omega_i$
to the present time and showed that the power-law term is the
dominant term the DGP model. To constrain the parameters of model we
fit our model with the new Gold sample and SNLS supernova data, CMB
shift parameter, position of the first and third peaks of power
spectrum of temperature fluctuations at the last scattering surface,
the Cluster Baryon Gas Mass Fraction, location of baryonic acoustic
oscillation peak observed by SDSS and large scale structure
formation data by $2$dFGRS. The best parameters obtained from
fitting with the new Gold sample data are: $h=0.62$,
$\Omega_m=0.28^{+0.03}_{-0.02}$,
$\Omega_{r_{c}}=0.13_{-0.01}^{+0.01}$ and
$\Omega_K=-0.002_{-0.053}^{+0.064}$ at $1\sigma$ confidence level
with $\chi^2_{min}/N_{d.o.f}=0.93$ and by using the SNLS data are:
 $\Omega_m=0.22^{+0.01}_{-0.01}$,
$\Omega_{r_{c}}=0.15_{-0.01}^{+0.01}$ and $\Omega_K=0.01_{-0.04
}^{+0.04}$ at $1\sigma$ confidence level with
$\chi^2_{min}/N_{d.o.f}=0.84$. Comparing our results to that of
previous results \cite{zong,maartens1} showed that large scale
structure observations from $2$dFGRS experiment had weak effect on
confining the acceptance intervals for the free parameters. The
observational constraint just by using SNIa+CMB indicated that our
universe is spatially open but combining these result with
SDSS+gas+LSS showed that our universe in the DGP model is very good
agreement with the spatially flat universe. In comparison between
$\Lambda$CDM and DGP in terms of $\chi^2_{\nu}$, Table \ref{tab7}
shows that these two models result almost same values.

\begin{table}[t]
\caption{The value of $\chi^2_{\nu}$ for $\Lambda$CDM and DGP
modified gravity models.} {\begin{tabular}{|c|c|c|} \hline
  Observation & $\Lambda$CDM &DGP    \\ \hline
  &&\\
SNIa (new Gold)& $0.92$ & $0.91$ \\
&&\\
 \hline

&&\\
SNIa(new Gold)+CMB &$0.93$&$0.94$ \\
 +SDSS& & \\
\hline
&&\\
SNIa(new Gold)+CMB & $0.93$&$0.93$ \\
 +SDSS+LSS& & \\ \hline
&&\\
 SNIa (SNLS)& $ 0.87$ & $0.85$ \\ &&\\ \hline
&&\\
SNIa(SNLS)+CMB & $0.86$&$0.85$ \\
 +SDSS& & \\
\hline
&&\\
SNIa(SNLS)+CMB & $0.85$&$0.84$ \\
 +SDSS+LSS& & \\
\hline
\end{tabular}
\label{tab7}}
\end{table}

 We also performed the age test,
comparing the age of old stars and old high redshift galaxies with
the age derived from this model. From the best fit parameters of
the model using new Gold sample and SNLS SNIa, we obtained an age
of $14.55_{-0.22}^{+0.32}$ Gyr and $13.88_{-0.15}^{+0.15}$ Gyr,
respectively, for the universe which is in agreement with the age
of old stars. We also chose two high redshift radio galaxies at
$z=1.55$ and $z=1.43$ with a quasar at $z=3.91$. The ages of the
two first objects were consistent with the age of universe,
i.e.,they were younger than the universe while the third one was
not.


\end{document}